\documentclass[12pt]{article}

\usepackage{latexsym}
\usepackage{amsmath}
\usepackage{multibox}
\usepackage{verbatim}
\usepackage{mathrsfs}

\usepackage{amssymb,amsmath}
\usepackage{amsthm}
\usepackage{amscd}
\usepackage{amssymb}
\usepackage{amsfonts}
\usepackage{url}
\usepackage{slashed}
\usepackage[latin1]{inputenc}
\usepackage[english]{babel}
\usepackage{putex}

 \csname
@addtoreset\endcsname{equation}{section}

\def\de{\partial}
\def\dsl{\not {\! \partial}}

\def\gz0{\gamma^{0}}

\def\nn{\nonumber}

\def\scs#1{\section{\sc #1}}
\def\scss#1{\subsection{\sc #1}}

\def\a{\alpha}
\def\b{\beta}

\def\d{\delta}

\def\e{\epsilon}

\def\l{\lambda}
\def\L{\Lambda}
\def\m{\mu}
\def\n{\nu}

\def\p{\pi}

\def\r{\rho}

\def\s{\sigma}

\def\vf{\varphi}

\newcommand{\beq}{\begin{equation}}
\newcommand{\eeq}[1]{\label{#1}\end{equation}}
\newcommand{\bea}{\begin{eqnarray}}
\newcommand{\eea}[1]{\label{#1}\end{eqnarray}}

\def\bs{\begin{split}}
\def\es{\end{split}}
\def\ba{\begin{array}}
\def\ea{\end{array}}
\def\bec{\begin{center}}
\def\ec{\end{center}}
\def\ba{\begin{align}}
\def\ena{\end{align}}

\def\12{\frac{1}{2}}

\def\pr{\partial}

\def\dag{\dagger}

\def\dsl{\not {\! \pr}}

\begin{document}

\begin{flushright}
{\today}
\end{flushright}

\vspace{25pt}

\begin{center}
{\Large\sc String Theory and The Velo--Zwanziger Problem}\\
\vspace{25pt}
{\sc Massimo~Porrati$^a$, Rakibur~Rahman$^b$, Augusto~Sagnotti$^b$}\\[15pt]
{\sl\small
$a$) Center for Cosmology and Particle Physics\\
    Department of Physics, New York University\\
    4 Washington Place, New York, NY 10003, USA\\
\vspace{6pt}
$b$) Scuola Normale Superiore and INFN\\
Piazza dei Cavalieri, 7\\I-56126 Pisa \ ITALY \\
\vspace{6pt}
e-mail: {\small \it
massimo.porrati@nyu.edu, rakibur.rahman@sns.it, sagnotti@sns.it}}\vspace{10pt}
\vspace{35pt}

{\sc\large Abstract}

\end{center}
We examine the behavior of the leading Regge trajectory of the
open bosonic string in a uniform electromagnetic background and present a consistent set of Fierz--Pauli
conditions for these symmetric tensors that generalizes the Argyres--Nappi spin-2 result.
These equations indicate that String Theory does bypass the Velo--Zwanziger problem,
\emph{i.e.} the loss of causality experienced by a massive high--spin field minimally coupled to
electromagnetism. Moreover, we provide some evidence that \emph{only} the first Regge
trajectory can be described in isolation and show that the open--string spectrum
is free of ghosts in weak constant backgrounds. Finally, we comment on the roles of
the critical dimension and of the gyromagnetic ratio.

\setcounter{page}{1}
\pagebreak
{\linespread{1.4}\tableofcontents}
\newpage

\scs{Introduction}\label{sec:intro}

In Quantum Field Theory, the available types of fundamental particles reflect the irreducible unitary
representations of the Poincar\'{e} group~\cite{wig1,wig2}, which exist for arbitrary (integer or
half--integer) values of the spin, not only for the handful of choices that underlie the Standard Model
of Electroweak and Strong Interactions or General Relativity. One is thus confronted with a
challenging problem, since higher--spin systems~\cite{solvay} are apparently fraught with grave difficulties,
so much so that in Minkowski space minimal interactions of \emph{massless} particles of
high spin with electromagnetism (EM) or gravity are not allowed~\cite{ad,ww,p}~\footnote{Minimal--like
interactions do become available for massive fields or in the presence of a cosmological constant~\cite{frvas},
but in both circumstances they ought to be regarded as byproducts of
higher--derivative ``seeds'', as recently stressed in~\cite{boulsund}.}. \emph{Massive} high--spin particles
certainly exist, in the form of hadronic resonances. Truly enough, these particles are
composite, so that the actual form factors describing their interactions are complicated functions of the
exchanged momenta. Still, in the quasi--collinear regime, when the exchanged momenta are small compared to
the particle masses, one expects that their dynamics is governed by consistent {\em local} actions. Moreover,
massive higher--spin modes play a role in (open) string spectra, where they describe excitations that are
generically unstable, and where the finite string size puts them again somewhat on the par with extended
composite systems. Some of the most spectacular novelties of String Theory \cite{String,polch}, however,
including (planar) duality, modular invariance and open--closed duality, rest heavily on their presence,
and this is by itself a compelling motivation to take a closer look at their properties.

Even if one restricts the attention to \emph{massive} higher--spin fields, a number of known actions
readily exhibit pathological behavior in the simplest possible settings, and in particular in constant
external backgrounds~\cite{vz,sham,deser}. A notorious example is provided by a charged
massive spin-2 field in a constant EM background in flat space, and it was indeed an early analysis of
this problem that led Fierz and Pauli~\cite{pf} to stress the importance of a Lagrangian formulation for
higher--spin systems. Their suggestion actually opened a wide avenue of research, with first complete
results in the 1970s \cite{sh,fronsdal} and new additions up to recent times~\cite{labastida,vas,bpt,fms},
but even the resulting Lagrangians, as we have anticipated, do not come to terms with the original problem.
Rather, in general they do not propagate the correct number of degrees of freedom (DoF) in the presence of
\emph{minimal} EM couplings, nor do they propagate their own DoFs only within the light cone. For spin $s=2$,
for instance, although the first difficulty can be overcome by an apparently unique choice of the gyromagnetic
ratio~\cite{fed}, $g=\12$\,, some of the modes suffer from lack of hyperbolicity or faster--than--light
propagation. This is the vexing ``Velo--Zwanziger problem''~\cite{vz}, which generally shows up when massive
charged fields with spin $s>1$ are minimally coupled to an EM background. The problem actually persists for
a wide class of non-minimal extensions, so that constructing consistent interactions for charged massive
higher--spin fields with EM from a field--theory vantage point appears to be a challenging task.

On the other hand, String Theory was originally meant to describe hadronic resonances, a plethora of
massive particles that, as we have already mentioned, typically carry high spins, and actually electric
charges as well, so that it should provide a valuable laboratory to investigate these exotic types of EM
interactions. And indeed, starting from the open bosonic string, Argyres and Nappi built long ago a
consistent Lagrangian~\cite{AN2} for a massive charged spin-2 field coupled to a constant EM background. In this case
the interactions reflect rather basic properties of String Theory, since they are induced by the EM deformation
of the free string developed in~\cite{Abouelsaood}. The resulting Lagrangian is nonetheless highly non-minimal,
but both its equations of motion (EoMs) and the constraints they give rise to are strikingly simple: they mimic
those of the free theory after some field redefinitions, which makes their consistency almost manifest.

The Lagrangian formulation attained via the BRST technique as in~\cite{AN2,AN1,string3} requires that the Fock
space be extended to include world--sheet (anti)ghosts. Hence, it involves in general a host of auxiliary fields,
and the procedure becomes rather cumbersome already for $s=3$~\cite{string3}. A result of this complication is
that it is not even clear, as of yet, whether the open bosonic string cures the Velo--Zwanziger problem
of its massive modes. And even if this were the case, a number of related questions still await a proper answer,
including the following two. Does consistency call for physical fields belonging to all Regge trajectories present
at a given string mass level, or could a (sub)leading Regge trajectory be consistent in isolation? Could one attain
a consistent description in non-critical dimensions as well? One would definitely like to arrive at a better understanding
of these issues, and to some extent we shall succeed. At the same time, while the BRST method gives gauge--invariant
Lagrangians for the modes of the open bosonic string in $d=26$, it is important to stress that gauge invariance alone
\emph{does not} guarantee that the resulting description be consistent, since after all any action can be made
gauge invariant via the St\"uckelberg formalism. In fact, the classical consistency of a dynamical system, and of the
Argyres--Nappi system in particular, rests on the behavior of the EoMs in a unitary gauge. Truly enough, a Lagrangian
formulation does guarantee that the resulting EoMs be algebraically consistent, but this key property can be also
verified directly, taking the EoMs themselves at face value.

In view of these considerations, we begin by formulating physical state conditions in the presence of a constant EM
background, without introducing any (anti)ghosts. These give rise to (partially gauge--fixed) EoMs that the string
fields must obey, and after removing some leftover modes that are pure gauge one can investigate directly their
consistency. One can work at any given mass level, because one is actually dealing with deformed free strings, and
in this fashion it is possible to identify particular sets of fields that are required for algebraic consistency. Given
these EoMs, one can also analyze explicitly both the actual propagating DoFs and their causal properties. The main
results of this paper are thus a relatively concise description of the consistent (non-minimal) EM interactions of
massive totally symmetric tensors of arbitrary spin that are present in String Theory and an explicit proof that
they provide a remedy for the Velo--Zwanziger problem, at least in $d=26$. More in detail, we show that any symmetric
tensor belonging to the first Regge trajectory of the open bosonic string can propagate independently, in a constant
EM background, the correct number of DoFs, and that these develop properly within the light cone. In addition,
we provide some evidence that fields belonging to subleading trajectories do not propagate consistently by themselves.
Let us emphasize, however, that our claims apply insofar as the EM field invariants, among which $F_{\m\n}\,F^{\m\n}$
is but one, are small in units of $m^2/e$, where $m$ is mass of the higher--spin field and $e$ is its electric charge.
This is an important qualification: if some invariant were ${\cal O}(1)$ in those units, a number of new phenomena
would present themselves, including Schwinger pair production~\cite{schwinger} and Nielsen--Olesen instabilities~\cite{nielsen}.
Their very existence implies precisely that any effective Lagrangian for a charged particle interacting with EM fields
can be reliable, even well below its own cutoff scale, only with this further proviso. The Velo--Zwanziger problem is
particularly important precisely because it appears well within the expected range of validity of the effective theory.

The paper is organized as follows. In Section~\ref{sec:open} we reconsider the world--sheet description of a charged
bosonic open string in a constant EM background and perform a careful analysis of the mode expansion and the Virasoro generators,
with emphasis on the behavior in the limit of vanishing total charge. Armed with this knowledge, in Sections \ref{sec:PSC}
and~\ref{sec:PSCEM} we translate the physical state conditions for string states into the language of string fields.
Section~\ref{sec:PSC} is actually devoted to free strings, but it is meant to make the reader better equipped for
understanding the more complicated case of charged strings, which we consider in Section~\ref{sec:PSCEM}, where we
show explicitly that String Theory indeed cures the Velo--Zwanziger problem for the symmetric tensors of the first
Regge trajectory. In Section~\ref{sec:NG} we present a
proof of the corresponding no--ghost theorem, showing that the Hilbert space of string states has a non-negative inner product
even in a (weak) constant EM background, which is crucial for consistency. In Section~\ref{sec:ANFed} we take
a closer look at spin-2 Lagrangians: in particular, Section~\ref{sec:AN} investigates the role of the critical dimension
in the consistency of the Argyres--Nappi construction, while Section~\ref{sec:linAN} elaborates on a possible route for
its generalization to arbitrary dimensions, and then solves a conundrum and clears up a misconception about the
gyromagnetic ratio of spin-2 particles. Finally, Section~\ref{sec:conclusion} contains some concluding remarks
and the two Appendices collect useful material on the massive $s=2$ system and on the bosonic string.

\vskip 36pt

\scs{Open strings in a constant EM background}\label{sec:open}

In this section we review in detail the mode expansion and the Virasoro algebra for a charged bosonic open
string in a constant EM background. The program originally started in~\cite{Abouelsaood} and has received
a wide attention in the literature, giving rise also to a number of applications (see, for instance,~\cite{reviews1,reviews2,Szabo},
for recent discussions). The novelty of our treatment in this section is a careful definition of the
expansion that makes it possible to reach smoothly the limits of neutral or free strings.

It will suffice to consider an open bosonic string whose endpoints lie on a space--filling D--brane.
A Maxwell field $A_\mu$ living in the world--volume of the D--brane couples to charges
$e_0$ and $e_\pi$ at the string endpoints, and this turns the string action into
\beq
S~=~\frac{1}{4\pi\alpha^{\prime}}\int d\tau d\sigma (\dot{X}_\mu\dot{X}
^\mu-X^{\prime}_\mu X^{\prime\mu})+\int d\tau d\sigma\,[e_0\delta(\sigma)+e_\pi\delta(\sigma
-\pi)]\,A_\mu(X)\dot{X}^\mu \ ,
\eeq{r1}
where the world sheet is chosen to be a strip of width
$\pi$ with a conformally flat metric of signature $(-,+)$ and $\alpha^{\prime}$ is
the string Regge slope. The $X^{\mu}$ are coordinates in the $d=26$ target
space, which we take to be Minkowski, while ``dot" and ``prime" denote derivatives with respect to
the world--sheet coordinates $\tau$ and $\sigma$.

In this paper we only consider electromagnetic backgrounds whose field strength
$F_{\mu\nu}$ is constant, so that one can choose the potential
\beq
A_\mu~=~- \ \tfrac{1}{2}\ F_{\mu\nu}\, X^\nu \ .
\eeq{gauge}
In units with $\alpha'=\tfrac{1}{2}$, the string sigma model then reads
\beq S=\frac{1}{2\pi}\int d\tau
d\sigma (\dot{X}_\mu\dot{X}^\mu-X^{\prime}_\mu X^{\prime\mu})+\frac{1}{2}\int d\tau d\sigma\,
[e_0\delta(\sigma)+e_\pi\delta(\sigma-\pi)]\,F_{\mu\nu}X^\mu\dot{X}^\nu\ . \eeq{r2}
As a result, the EoMs are those of the usual free string,
\beq \ddot{X}_\mu-X^{\prime\prime}_\mu=0 \ , \eeq{r4}
while the boundary conditions are affected by the new terms and become
\bea X^\prime_\mu&=&-\pi e_0F_{\mu\nu}\dot{X}^\nu \qquad (\sigma=0)\ ,\label{r5}\\
X^\prime_\mu&=&+\pi e_\pi F_{\mu\nu}\dot{X}^\nu \qquad (\sigma=\pi)\ .\eea{r6}
Let us also define, for later use,
\beq
e~\equiv~e_0\,+\, e_\pi \ . \eeq{etot}
\vskip 24pt

\scss{Mode expansion}\label{sec:mode}

In solving this boundary value problem, one should take into account that the $F_{\mu\nu}\rightarrow0$ limit
ought to recover the \emph{free} string mode expansion, which is recalled in Appendix~\ref{sec:states} along
with some useful facts about the free mode functions. We denote with $\mathbb{N}_0$ ($\mathbb{N}_1$) the set of all
natural numbers including (excluding) $0$, and we adopt for matrix multiplications a concise notation, so that $A^{\mu\nu}
u_\nu=A^\mu_{~\nu}u^\nu\equiv(Au)^\mu,~u_\mu A^{\mu\nu}=u^\mu A_\mu^{~\nu} \equiv(uA)^\nu,~A^{\mu\rho}B_\rho^{~\nu}=
A^\mu_{~\rho}B^{\rho\nu}\equiv(AB)^{\mu\nu}$, and we write $\eta^{\mu\nu}$ as $\mathbf{1}^{\mu\nu}$, and
$\delta^\mu_\nu$ as $\mathbf{1}^\mu_\nu$\,.

We can thus present the solution \cite{Abouelsaood} of Eqs.~(\ref{r4})--(\ref{r6}) in the form
\bea X^\mu(\tau,\sigma)&=&x^\mu+\left[\left(\frac{\text
{e}^{-G_0}}{2}\cdot\frac{\text{e}^{G(\tau+\sigma)}-M_+}{G}\;+\;\frac{\text{e}^{+G_0}}{2}\cdot\frac
{\text{e}^{G(\tau-\sigma)}-M_-}{G}\right)\alpha_0\right]^\mu\nonumber\\&&+\,\frac{i}{2}\sum_{m\neq0}
\left[\left(\tfrac{1}{m\mathbf{1}+iG}\right)\left\{\text{e}^{-i(m\mathbf{1}+iG)(\tau+\sigma)-G_0}+\text{e}^{-i(m\mathbf{1}+iG)
(\tau-\sigma)+G_0}\right\}\alpha_m\right]^\mu, \eea{r12}
where the matrices
\beq G_0~=~\tanh^{-1}(\pi e_0F)\ ,\qquad G_\p~=~\tanh^{-1}(\pi e_\p F)\ ,
\qquad G~=~\frac{1}{\pi}\, [\,G_0+G_\pi\,]\ ,\eeq{r13}
are uniquely determined by the boundary conditions~(\ref{r5}) and (\ref{r6}), while the matrices $M_\pm$ are
additional functions of $F$ whose forms will be specified shortly. No ambiguities are met in these
expressions, since $G, G_0, M_\pm$ and their inverses are all functions of $F$ only, and are thus
mutually commuting. Note, finally, that the matrices $(m\mathbf{1}\pm iG)$, with $m\neq0$, are always
invertible whenever the EM field invariants are sufficiently small.

In writing the mode expansion~(\ref{r12}) we required that, in the $F\rightarrow0$ limit, the $\alpha^\mu_m$
reduce for any given $m$ to the modes of the free string, so that the same must hold true for the mode function
matrices. This is readily seen to be the case for the ``oscillator'' modes in the second line of Eq.~(\ref{r12}),
since $G_0$ and $G$ tend to zero in this limit. On the other hand, the requirement that the coefficient matrix of
$\alpha^\mu_0$ reduces, in the limit, to $\bar{\Psi}_0=\tau$, poses on $ M_\pm$ the non-trivial condition that
\beq M_\pm\,=\,\mathbf{1}\, \pm\, \gamma\, G\ +\ \mathcal{O}(G^2)\ .\eeq{r13.5}
As we shall see shortly, the constant $\gamma$, along with the whole $\mathcal{O}(G^2)$ term, can be completely
determined requiring that the $x^\mu$'s in~(\ref{r12}) be \emph{standard commuting center--of--mass coordinates}.
This choice will also lead to a smooth limit of the resulting expressions in the dipole ($e_0+e_\pi=0$) or free
($e_0=e_\pi=0$) cases, contrary to some claims that have appeared in~\cite{Szabo}.

Now notice that the matrix--valued functions (of $\tau$ and $\sigma$),
\beq \Psi'_m
(\tau,\sigma)~=~\tfrac{i/2}{\sqrt{m\mathbf{1}+iG}}\ \text{e}^{-i(m\mathbf{1}+iG)\tau}[\,\text{e}^
{-i(m\mathbf{1}+iG)\sigma-G_0}+\text{e}^{i(m\mathbf{1}+iG)\sigma+G_0}\,]\qquad m\in\mathbb{N}_0 \ , \eeq{r14}
form an orthonormal set, since
\beq (\Psi'_m,\Psi'_n)~\equiv~\frac{1}{\pi}\int_0^\pi d\sigma \Psi_m^{\prime\dag}
(\tau,\sigma)\star\Psi'_n(\tau,\sigma)\,=\,\delta_{mn}\qquad m,n\in\mathbb{N}_0 \ , \eeq{r15}
if $\star$ is defined as
\beq \star~\equiv~i\overleftrightarrow{\partial_\tau}-i\tanh G_0\,\delta(\sigma)-i\tanh G_\pi\,
\delta(\sigma-\pi)\ .\eeq{r16}
A constant has thus a non-vanishing norm, and moreover it is orthogonal to all other functions in
Eq.~(\ref{r14}),
\beq (1,1)~=~-ieF\ ,\qquad(1,\Psi'_m)~=~0 \qquad m\in\mathbb{N}_0\ ,\eeq{r16.5}
where $e$ is the total string charge defined in Eq.~\eqref{etot}.
Therefore, in view of the mode expansion (\ref{r12}) one can write
\beq \alpha_m^\mu\,=\,(\sqrt{m\mathbf{1}+iG})^\mu_{~\nu}
(\Psi'_m, X)^\nu\qquad m\in\mathbb{N}_0 \ ,\eeq{r17}
taking into account the reality of $X^\mu$ and the relations
\beq (\Psi^{\prime*}_m,\Psi^{\prime*}_n)\,=\,\delta_{mn},\qquad (\Psi^{\prime}_m,\Psi^{\prime*}_n)\,=\,
-(\Psi^{\prime*}_m,\Psi^{\prime}_n)\,=\,i\delta_{m0}\qquad m,n\in\mathbb{N}_0\ ,\eeq{r17.1}
and one can let
\beq \alpha_{-m}^\mu\,=\,(\sqrt{m\mathbf{1}-iG})^\mu_{~\nu}(\Psi^{\prime*}_m, X)^\nu\qquad m\in\mathbb{N}_0 \ .
\eeq{r17.5}

Upon quantization, the string modes $\alpha_{m}^\mu$ with $m\in\mathbb{Z}$ of Eqs.~(\ref{r17}) and (\ref{r17.5})
become operators that obey non-trivial commutation relations. In order to quantize the system, one can first note
that the canonical momentum for the sigma model~(\ref{r2}) is
\beq P^\mu(\tau,\sigma)\,=\,
\frac{1}{\pi}\,\dot{X}^\mu(\tau,\sigma)\,-\,\frac{1}{2}\, [\,e_0\delta(\sigma)+e_\pi\delta(\sigma-\pi)\,]
\,F^{\mu\nu}X_\nu(\tau,\sigma)\ , \eeq{r18}
and then require that $X$ and $P$ satisfy the equal time commutation relations
\bea [X^\mu(\tau,\sigma),P^\nu(\tau,\sigma')]&=&i\eta^{\mu\nu}
\delta(\sigma-\sigma')\ ,\label{r19}\\\,[X^\mu(\tau,\sigma),X^\nu(\tau,\sigma')]&=&[P^\mu(\tau,\sigma),P^\nu
(\tau,\sigma')\,]~=~0\ .\eea{r20}
Using Eqs.~(\ref{r17}), (\ref{r17.5}) and (\ref{r18})--(\ref{r20}) it is then simple to show that
\beq [\,\alpha_m^
\mu,\alpha_n^\nu\,]\,=\,(m\, \eta^{\mu\nu}\,+\,i\,G^{\mu\nu})\,\delta_{m,-n}\qquad m,n\in\mathbb{Z}\ .\eeq{r22}

Before
computing the other commutators, let us elaborate on the meaning of these results. In physically interesting
situations, away from instabilities, the matrices $\sqrt{m\mathbf{1}\pm iG}$ are always invertible when $m\in\mathbb{N}_1$
so that, on account of Eq.~(\ref{r22}),
\beq a_m^\mu~\equiv~\left[\frac{1}{\sqrt{m\mathbf{1}+iG}}\ \alpha_m\right]^\mu,\qquad
a_m^{\dag\mu}~\equiv~\left[\frac{1}{\sqrt{m\mathbf{1}-iG}}\ \alpha_{-m}\right]^\mu\qquad m\in\mathbb{N}_1\eeq{oscillators}
are an infinite set of creation and annihilation operators:
\beq [\,a_m^\mu,a_n^{\dag\nu}\,]
\,=\,\eta^{\mu\nu}\delta_{mn},\qquad [\,a_m^\mu,a_n^{\nu}\,]\,=\,[\,a_m^{\dag\mu},a_n^{\dag\nu}\,]\,=\,0\qquad
m,n\in\mathbb{N}_1 \ .\eeq{r23}
When $m=0$, however, one cannot reach this point
starting from~(\ref{r22}), since $\sqrt{G}$ is not invertible when $F\neq 0$ in some Lorentz frame (and obviously so
when $eF=0$). The $\alpha_0^\mu=\alpha_0^{*\mu}$, on the other hand, are well--defined, and their commutation relations
read
\beq [\,\alpha_0^\mu,\alpha_0^\nu\,]~=~i\, G^{\mu\nu}\ .\eeq{r24}
Naively, one would expect that $\alpha_0^\mu$ play the role of a covariant momentum, since after all
it reduces to the string momentum $\bar{\alpha}_0^\mu=p^\mu$ when $F$ vanishes. Furthermore, the string
Hamiltonian is
\beq H~\equiv~\int_0^\pi d\sigma[\,P^\mu\dot{X}_\mu-\mathcal{L}\,]~=~
\frac{1}{2\pi}\int_0^\pi d\sigma(\dot{X}^2+X'^2)~=~\tfrac{1}{2}\sum_{m\in\mathbb{Z}}\alpha_{-m}\cdot\alpha_m \ ,
\eeq{r25}
where the last equality is due to a consequence of the mode expansion~(\ref{r12}), namely
\beq (\dot{X}\pm X')^\mu(\tau,\sigma)~=~\sum_{m\in\mathbb{Z}}\left[\text{e}^{-i(m\mathbf{1}+iG)(\tau\pm\sigma)
\mp G_0}\right]^\mu_{~\nu}\alpha_m^\nu\qquad \sigma\in[0,\pi]\ .\eeq{r26}
In view of Eqs.~(\ref{r23}) and (\ref{r25}), therefore,
\beq H~=~\tfrac{1}{2}\ \alpha_0^2+(\text{Stringy Oscillator Contributions})\ ,\eeq{r27}
whereas for a charged point particle the Hamiltonian would read
\beq
H~=~\tfrac{1}{2}\, p^2_{\text{cov}}\ , \eeq{hamiltonian}
where $p^\mu_{\text{cov}}$ is the covariant momentum. This vindicates the identification of $\alpha_0^\mu$
with the covariant momentum, up to a matrix redefinition that is needed since
\beq [\,p^\mu_{\text{cov}},p^\nu_{\text{cov}}\,]~=~ieF^{\mu\nu}\ .\eeq{r28}
Comparing Eqs.~(\ref{r24}) and (\ref{r28}), one is thus led to conclude that
\beq \alpha_0^\mu~=~Q^\mu_{~\nu}\,p^\nu_{\text{cov}} \ ,
\qquad QQ^T~=~\frac{G}{eF} \ ,\eeq{r29}
so that the covariant derivative,
\beq
D^\mu~\equiv~ip^\mu_{\text{cov}}\ ,\eeq{covder}
finally obeys the desired commutation relation:
\beq [\,D^\mu,D^\nu\,]~=~-ieF^{\mu\nu}\ .\eeq{commutator}
Notice that the matrix $Q$ reduces to unity in the $F\rightarrow0$ limit, and that one can choose it to be symmetric, letting
\beq Q~=~\sqrt{\frac{G}{eF}} \ .\eeq{r30}
For future reference, it is actually convenient to define a different covariant derivative,
\beq \mathcal D^\m~\equiv~\left(\sqrt{G/eF}\right)^{\m\n}D_\n~=~i\,\a_0^\m~,\eeq{fancyD}
such that
\beq
\qquad [\mathcal D^\m,\mathcal D^\n]~=~-\, i\,G^{\m\n} \ .
\eeq{fancyD2}

Now one can easily compute the commutators $[x^\mu,\alpha_m^\nu]$, starting from the relation
\beq [X^\mu(\tau,\sigma),
(\dot{X}+X')^\nu(\tau,\sigma')]~=~[X^\mu(\tau,\sigma),\dot{X}^\nu(\tau,\sigma')]~=~i\pi\eta^{\mu\nu}\delta
(\sigma-\sigma')\ , \eeq{r31}
and from Eqs.~(\ref{r12}) and (\ref{r26}), with the end result
 \beq [\,x^\mu,
\alpha_m^\nu\,]~=~\frac{i}{2}\left[\,\text{e}^{-G_0}M_+ + \text{e}^{G_0}M_-\,\right]^{\mu\nu}\delta_{m0}\ .
\eeq{r32}
The identification of $x^\mu$ as center--of--mass coordinates now demands that
\beq
[x^\mu,p_{\text{cov}}^\nu]~=~i\eta^{\mu\nu}\ ,\eeq{}
and in view of Eqs.~(\ref{r29}), (\ref{r30}) and (\ref{r32}) this leads to
\beq \text{e}^{-G_0}M_+ + \text{e}^{G_0}M_-~=~2\,\sqrt{\frac{G}{eF}}~.\eeq{r33}
Finally, starting from the fundamental commutator
\beq
[X^\mu(\tau,\sigma),X^\nu(\tau,\sigma')]~=~0\ ,
\eeq{etcr0}
and making use of the mode expansion~(\ref{r12}) and of Eqs.~(\ref{r22}) and (\ref{r32}), one can see that
\beq [\,x^\mu,x^\nu\,]\,=\,0\ ,\eeq{r34}
so that the center--of--mass coordinates are indeed mutually commuting, on account of Eq.~(\ref{r33}).

In order to find $M_\pm$ explicitly, one can now demand that the mode expansion~(\ref{r12}) be symmetric under the
flip operation $\sigma\rightarrow
(\pi-\sigma)$, to be combined with the interchange of the charges at the endpoints, $e_{0,\pi}\rightarrow e_{\pi,0}$, and
with the flip properties of the oscillators, $\alpha_m\rightarrow(-1)^m\alpha_m$. In view of~(\ref{r13.5})
it is clear that the $M_\pm$ do not transform under such a flip. Therefore, Eq.~(\ref{r33}) gives
\beq \text{e}^
{-G_\pi}M_+ + \text{e}^{G_\pi}M_-~=~2\,\sqrt{\frac{G}{eF}}\ , \eeq{r33.5}
and thus Eqs.~(\ref{r33}) and (\ref{r33.5}) finally lead to
\beq M_\pm~=~\sqrt{\frac{G}{eF}}\;\text{sech}\left[\,\tfrac{1}{2}(G_\pi-G_0)\right]
\,\text{e}^{\pm\pi G/2}~=~\mathbf{1}\pm\tfrac{1}{2}\pi G+\mathcal{O}(G^2)\ .\eeq{r35}

Let us note that, if the string is a dipole, so that $e_0=-e_\pi$, $G$ vanishes but $G_0$ remains finite,
while the covariant momentum reduces to the ordinary momentum. From Eqs.~(\ref{r29}), (\ref{r30}) and (\ref{r35}),
one can see that the mode expansion~(\ref{r12}) of the charged string reduces smoothly to a corresponding expression
for the neutral one,
\bea X^\mu_{\text{neut}} (\tau,\sigma)&=&x^\mu+\left[\frac{\tau\, -\,(\sigma-\pi/2)\, \pi e_0\,F}{\sqrt{\mathbf{1}
-(\pi e_0F)^2}}\right]^{\mu\nu}\,\left[\,\frac{1}{\sqrt{\mathbf{1}-(\pi e_0F)^2}}\right]_{\nu\sigma}p^\sigma\nonumber\\
&+&\frac{i}{2}\sum_{m\neq0}\tfrac{1}{m}\left[\text{e}^{-im(\tau+\sigma)-G_0}+\text{e}^{-im(\tau-\sigma)+G_0}\,\right]^\mu_{~\nu}\
\alpha_m^\nu \ .\eea{neut1}
It is important to notice that this expression differs from Eq.~(2.26) of~\cite{Szabo}, and moreover that the charged
string mode functions are not quite given by~(\ref{r14}). Rather, they are
\bea \Psi_m(\tau,\sigma)&=&\tfrac{i/2}{\sqrt{m\mathbf{1}+iG}}\,\text{e}^{-i(m\mathbf{1}+iG)\tau}\left[\,\text{e}^
{-i(m\mathbf{1}+iG)\sigma-G_0}+\text{e}^{i(m\mathbf{1}+iG)\sigma+G_0}\,\right]\qquad m\in\mathbb{N}_1\ ,\label{r36}\\
\Psi_0(\tau,\sigma)&=&\frac{\text{e}^{-G_0}}{2}\cdot\frac{\text{e}^{G(\tau+\sigma)}-M_+}{G}\;+\;\frac{\text{e}^
{+G_0}}{2}\cdot\frac{\text{e}^{G(\tau-\sigma)}-M_-}{G} \ ,\eea{r37}
together, of course, with a constant mode. While
$\Psi_m=\Psi_m'$ for all $m\in\mathbb{N}_1$, $\Psi_0$ is in fact a linear combination of $\Psi'_0$ and the
constant mode. The latter fact is crucial, in that it guarantees a smooth $F\rightarrow0$ limit. The inner
product is still defined according to Eqs.~(\ref{r15}) and (\ref{r16}), so that the following orthogonality relations
for $m,n\in\mathbb{N}_0$ hold:
\beq (\Psi_m,\Psi_n)~=~\delta_{mn}(1-\delta_{m0})\ ,\qquad (1,1)~=~-ieF\ ,\qquad
(1,\Psi_m)~=~i\sqrt{eF/G}\,\delta_{m0} \ .\eeq{r38}

These mode functions naturally split once more into two
mutually orthogonal subsets, particle--like $\{1,\Psi_0\}$, and string--like $\{\Psi_{m\in\mathbb{N}_1}\}$, and the
infinitely many string--like modes form an orthonormal set of functions with respect to the inner product of
Eq.~\eqref{r15}. The two particle--like modes have a non-vanishing
inner product, and the norm of $\Psi_0$ vanishes while that of $1$ is $\mathcal{O}(F)$. Everything thus parallels
the relations~(\ref{r10}) and (\ref{r11}) for free strings, because of the particular linear
combination appearing in~(\ref{r37}), and in the present case Eq.~(\ref{r11.5}) generalizes to
\beq x^\mu\,=\, i(\Psi_0, X)^\mu\ ,\qquad
p_{\text{cov}}^\mu~=~-i\,(\mathbf{1}+eF\Psi_0, X)^\mu\ .\eeq{r39}

\vskip 24pt

\scss{Virasoro generators}\label{sec:Virasoro}

As we have seen already, the world--sheet action under consideration differs from the free string Polyakov action only
by boundary terms. The latter, however, do not depend on the world--sheet metric, since they are obtained via the
pullback of target--space one--forms. Therefore, the constraints that are to be imposed after gauge fixing take the
same form as in the free theory:
\beq (\dot{X}\pm X')^2~=~0 \ .
\eeq{r40}
With the Virasoro generators $L_n$ defined as
\beq \sum_{n\in\mathbb{Z}}L_n\, \text{e}
^{-in(\tau\pm\sigma)}~\equiv~\tfrac{1}{2}(\dot{X}\pm X')^2(\tau,\sigma) \ ,\eeq{r41}
one can formally extend the range of $\sigma$ to $[-\pi,\pi]$ with the help of Eq.~(\ref{r26}), to write
\beq L_n~=~\frac{1}
{4\pi}\int_{-\pi}^{\pi}d\sigma\, \text{e}^{in(\tau+\sigma)}(\dot{X}\pm X')^2(\tau,\sigma)~=~\tfrac{1}{2}
\sum_{m\in\mathbb{Z}}\alpha_{n-m}\cdot\alpha_m\ .\eeq{r42}
The final expression is identical to that of the free theory, but as we have seen the $\alpha_m$'s now have the
commutation relations~(\ref{r22}). One can work out the commutation relations obeyed by the Virasoro generators,
paying attention as usual to the central extension. The end result is the emergence, in the constant EM background,
of an additive contribution to $L_0$, so that
\beq L_0~\rightarrow~L_0+\tfrac{1}{4}\text{Tr}G^2 \ , \eeq{r52}
but up to this shift the Virasoro algebra,
\beq [L_m,L_n]~=~(m-n)L_{m+n}\,+\,\tfrac{1}{12}\,d(m^3-m)\delta_{m,-n}\ , \eeq{r53}
remains precisely as in the free theory. The shift, however, has an important effect,
since it reflects itself in deformed masses for the open--string excitations.
\vskip 36pt

\scs{Physical state conditions for free strings}\label{sec:PSC}

Having spelled out the key properties of the (charged) open string modes, the corresponding Virasoro generators
and their algebra, we can now turn to the physical state conditions for (charged) string states. In this section
we actually begin by considering free strings, as this provides some valuable insights for the more complicated
study of charged strings, which will be the subject of the next section.

Let us begin by recalling that a string state $|\Phi\rangle$ is called ``physical'' if it satisfies the conditions~(see
\emph{e.g.}~\cite{String,polch})
\beq (L_n-\delta_{n0})|\Phi\rangle~=~0\qquad n\in\mathbb{N}_0\ .\eeq{p10}
Actually, in view of the Virasoro algebra, it suffices to demand that
\bea (L_0-1)|\Phi\rangle&=&0\ ,\label{p11}\\ L_1|\Phi\rangle
&=&0\ ,\label{p12}\\L_2|\Phi\rangle&=&0 \ .\eea{p13}
Once Eqs.~(\ref{p9.1})--(\ref{p9.3}) are used, for the symmetric tensors of the leading Regge trajectory these physical state conditions translate
into the well--known Fierz--Pauli conditions, namely the Klein--Gordon equation and the conditions
that their divergences and traces vanish.

\vskip 24pt

\scss{Massless level: $N=1$}

The generic state at this level is
\beq |\Phi\rangle~=~A_\mu(x)\,a^{\dag\mu}_1\,|0\rangle \ .\eeq{p16}
In this case Eq.~(\ref{p13}) is empty, and one thus obtains the Maxwell equations in the Lorenz gauge,
\beq \Box A_\mu~=~0\ ,\qquad \partial^\mu
A_\mu~=~0 \ , \eeq{p19}
for the massless vector field $A_\mu$. Notice that these equations are invariant
under the \emph{on--shell} gauge transformation
\beq
\d A_\m \, = \, \partial_\m \, \alpha \ ,\qquad \Box \, \a \, = \, 0 \ . \eeq{p19.5}
\vskip 24pt

\scss{First massive level: $N=2$}

In this case a generic state takes the form
\beq |\Phi\rangle~=~h_{\mu\nu}(x)\,a^{\dag\mu}_{1}a^{\dag\nu}_{1}\,
|0\rangle~+~\sqrt{2}\,i\,B_\mu(x)\,a^{\dag\mu}_{2}\,|0\rangle \ , \eeq{p20}
and Eqs.~(\ref{p11})--(\ref{p13}) give
\bea &(\Box-2)h_{\mu\nu}~=~0\ ,\qquad (\Box-2)B_\mu~=~0\ ,&\label{p24}\\&\partial^\mu
h_{\mu\nu}-B_\nu~=~0\ ,\qquad h^\mu_{~\mu}+2\partial^\mu B_\mu~=~0\ .&\eea{p25}
One can verify that these equations possess, in an arbitrary space--time dimension $d$, the \emph{on--shell}
gauge symmetry
\bea \delta h_{\mu\nu}&=&(\partial_\mu\xi_\nu+\partial_\nu\xi_\mu)-\left(\tfrac{10}{d+4}\right)
\eta_{\mu\nu}\, \partial\cdot\xi\ ,\label{p27}\\ \delta B_\mu&=&2\xi_\mu+\left(\tfrac{d-6}{d+4}\right)
\partial_\mu\, \partial\cdot\xi \ ,\eea{p28}
where the gauge parameter $\xi_\mu$ satisfies the condition
\beq (\Box-2)\, \xi_\mu\, = \, 0\ . \eeq{p26}

One could have arrived at the string field equations~(\ref{p24}) and~(\ref{p25}) via the BRST construction,
following \cite{ko}. There would be more fields to begin with than those present in~(\ref{p20}), but one could
(partially) gauge fix the EoMs using the BRST symmetry, which holds only in the critical dimension,
to finally recover Eqs.~\eqref{p24} and~\eqref{p25}. The gauge symmetry~(\ref{p27})--(\ref{p26}) actually
holds in an arbitrary number of dimensions, and is thus more general than what the BRST method would give. In fact,
proceeding from a field theory perspective, one can also build rather naturally, at least for the first few mass levels,
a generalization of the BRST symmetry that is available in an arbitrary number of space--time dimensions \cite{RS}.

One can now gauge away the vector field $B_\mu$, in any number of space--time dimensions, making use of the gauge
parameter $\xi_\mu$, and thus end up with the spin-2 Fierz--Pauli system
\beq (\Box-2)h_{\mu\nu}~=~0\ ,\qquad \partial^\mu h_{\mu\nu}~=~0\ ,\qquad h^\mu_{~\mu}~=~0\ ,\eeq{p29}
so that $h_{\mu\nu}$ is a massive spin-2 field, with $(\text{mass})^2=2$ (or $1/\alpha'$ taking into account
our choice of units), that obeys the
Fierz--Pauli conditions.

It is also possible to arrive at~(\ref{p29}) by a gauge--fixing procedure that does not involve solving a differential
equation for the gauge parameter. To this end, one can define a new vector field $B'_\m$ whose
gauge variation is algebraic in $\xi_\m$\,,
\beq B'_\m~\equiv~B_\m +\left(\tfrac{d-6}{10}\right)\left[\de^{\n}
h_{\m\n}+\tfrac{1}{8}\left(\tfrac{d+4}{d-1}\right)\de_\m h^\n_{~\n}\right]~,\qquad \delta B'_\m~=~\left(\tfrac{d+4}{5}\right)\xi_\m\ .\eeq{g1}
It is now evidently possible to choose $B'_\m$ in such a way that $B_\m=0$, whence Eq.~(\ref{p29}) follows.

\vskip 24pt

\scss{Second massive level: $N=3$}

A generic state at this mass level is
\beq |\Phi\rangle~=~\phi_{\mu\nu\rho}(x)\,a^{\dag\mu}_{1}a^{\dag\nu}_{1}
a^{\dag\rho}_{1}\,|0\rangle~+~\tfrac{i}{\sqrt2}\,[\,h_{\mu\nu}(x)+A_{\mu\nu}(x)\,]\,a^{\dag\mu}_{2}a^{\dag\nu}_{1}\,|0\rangle
-\tfrac{1}{\sqrt3}\,B_\mu(x)\,a^{\dag\mu}_{3}\,|0\rangle \ , \eeq{p30}
where $A_{\mu\nu}$ is an antisymmetric tensor while both $\phi_{\mu\nu\rho}$ and $h_{\mu\nu}$ are symmetric tensors.
Eqs.~(\ref{p11})--(\ref{p13}) give rise to
\bea &(\Box-4)\phi_{\mu\nu\rho}~=~0\ ,\qquad (\Box-4)[h_{\mu\nu}+A_{\mu\nu}]~=~0\ ,\qquad (\Box-4)B_\mu~=~0\ ,&
\label{p34}\\& 3\partial^\mu\phi_{\mu\nu\rho}-h_{\nu\rho}~=~0\ ,\qquad\partial^\mu A_{\mu\nu}-\partial^\mu h_{\mu\nu}+2B_\nu~=~0\ ,&\label{p35}\\&3\phi^\mu_{~\mu\nu}+\partial^\mu h_{\mu\nu}+\partial^\mu A_{\mu\nu}
-B_\nu~=~0\ ,&\eea{p36}
and in the critical dimension, $d=26$, this system is invariant under the \emph{on--shell} gauge transformations
\bea \delta\phi_{\mu\nu\rho}&=&\partial_{(\mu}\lambda_{\nu\rho)}+\eta_{(\mu\nu}\xi_{\rho)}~,\label{p37}\\
\delta h_{\mu\nu}&=&12\lambda_{\mu\nu}+3\eta_{\mu\nu}\partial\cdot\xi+3\partial_{(\mu}\xi_{\nu)}+3\partial
_{(\mu}(\partial\cdot\lambda)_{\nu)}~,\label{p38}\\\delta A_{\mu\nu}&=&-15\partial_{[\mu}\xi_{\nu]}-3\partial
_{[\mu}(\partial\cdot\lambda)_{\nu]}~,\label{p39}\\\delta B_\mu&=&36\xi_\mu+18(\partial\cdot\lambda)_\mu
-\tfrac{9}{2}\partial_\mu\partial\cdot\xi~,\eea{p40}
where $\lambda_{\mu\nu}=\lambda_{\nu\mu}$ and the gauge
parameters are subject to the conditions
\bea &(\Box-4)\lambda_{\mu\nu}~=~0~,\qquad (\Box-4)\xi_\mu~=~0~,&\label{p41}\\
&\lambda^\mu_{~\mu}~=~-2\partial\cdot(\partial\cdot\lambda)-\tfrac{17}{2}\partial\cdot\xi~.&\eea{p42}
Just as in the preceding $N=2$ case, here it should be possible to render the gauge symmetry valid for arbitrary
values of $d$ by judicious modifications of the coefficients appearing in the gauge transformations and in the trace
constraint~(\ref{p42}). We hope to return to this point for the complete spectrum in a future publication \cite{RS}.

It is possible to gauge away the vector field $B_\mu$ using the parameter $\xi_\mu$. On the other hand, because of
the trace constraint~(\ref{p42}) on the gauge parameter $\lambda_{\mu\nu}$, one can only set to zero the traceless
part of $h_{\mu\nu}$. This gauge fixing thus reduces the system to \footnote{In this step, one can forego
the need to solve differential equations for the gauge parameters, as we have already seen for the $N=2$ case.}
\bea &(\Box-4)\phi_{\mu\nu\rho}~=~0\ ,\qquad (\Box-4)A_{\mu\nu}~=~0\ ,\qquad (\Box-4)h~=~0\ ,&\label{p43}\\
& 3\partial^\mu\phi_{\mu\nu\rho}~=~\eta_{\nu\rho}h \ ,\qquad\partial^\mu A_{\mu\nu}~=~\partial_\nu h\ ,\qquad
3\phi^\mu_{~\mu\nu}+\partial^\mu A_{\mu\nu}~=~-\partial_\nu h\ ,&\eea{p44}
where $h^\mu_{~\mu}\equiv d\,h$\,. One can now show that $h$ is an auxiliary scalar field: as is well known, it is
essential for writing a local Lagrangian for a massive spin-3 field, but it is set to zero on--shell if
Eqs.~\eqref{p43} and \eqref{p44} are combined with their traces and divergences. Once $h$ is eliminated,
the system reduces to
\bea &(\Box-4)\phi_{\mu\nu\rho}~=~0\ ,\qquad \partial^\mu\phi_
{\mu\nu\rho}~=~0\ ,\qquad \phi^\mu_{~\mu\nu}~=~0\ .&\label{p46}\\&(\Box-4)A_{\mu\nu}~=~0 \ ,\qquad \partial^\mu
A_{\mu\nu}~=~0\ .&\eea{p47}
Eq.~(\ref{p46}) are the Fierz--Pauli conditions for a massive spin-3 symmetric tensor field, $\phi_{\mu\nu\rho}$,
that belongs to the first Regge trajectory for the open bosonic string and whose squared mass is $2/\alpha'$~(taking into
account our choice of units), while Eq.~(\ref{p47}) are the corresponding equations for an antisymmetric rank-2 field of the same
mass \footnote{In $d=4$, $A_{\mu\nu}$ is equivalent to a massive vector field, and Eq.~(\ref{p47}) is clearly
consistent with this fact.}. These are the two physical fields that appear at the second massive level of the open
bosonic string, and in the free theory their EoMs are completely decoupled.

\vskip 36pt

\scs{Physical state conditions in an EM background}\label{sec:PSCEM}

The construction of a generic string state in the presence of a constant EM background follows the same
lines as in the free theory, because the two cases rest on identical sets of creation and annihilation
operators. While the Virasoro generators themselves are different, their algebras are the same in both
cases, so that the physical state conditions still read
\bea (L_0\, - \, 1)|\Phi\rangle&=&0 \ ,\label{b11}\\ L_1|\Phi\rangle&=&0\ ,\label{b12}\\L_2|\Phi\rangle&=&0 \ ,\eea{b13}
where now
\bea L_0&=&-\tfrac{1}{2}\mathcal{D}^2+\sum_{m=1}^{\infty}(m+iG)_{\mu\nu}\,a_m^{\dag\mu}a_m^\nu+
\tfrac{1}{4}\,\text{Tr}G^2\nonumber\\&\equiv&-\tfrac{1}{2}\mathcal{D}^2+\left(\mathcal{N}+\tfrac{1}{4}\,
\text{Tr}G^2\right)+i\sum_{m=1}^{\infty}G_{\mu\nu}\,a_m^{\dag\mu}a_m^\nu~,\label{b9.1}\\L_1&=&-i
\left[\sqrt{1+iG}\right]_{\mu\nu}\mathcal{D}^\mu a_1^\nu+\sum_{m=2}^{\infty}\left[\sqrt{(m+iG)
(m-1+iG)}\right]_{\mu\nu}a_{m-1}^{\dag\mu}a_m^\nu~,\eea{b9.2} \bea L_2~~=~~-i\left[\sqrt{2+iG}\right]
_{\mu\nu}\mathcal{D}^\mu a_2^\nu&+&\tfrac{1}{2}\left[\sqrt{1+G^2}\right]_{\mu\nu}a_1^\mu a_1^\nu
\nonumber\\&+&\sum_{m=3}^{\infty}\left[\sqrt{(m+iG)(m-2+iG)}\right]_{\mu\nu}a_{m-2}^{\dag\mu}
a_m^\nu~,\eea{b9.3}
and $\mathcal D_\m$ was defined in Eq.~(\ref{fancyD}). Notice that here we are going to define the number
operator $\mathcal{N}$ in such a way that its eigenvalues $N$ are integers, just as for free strings:
\beq \mathcal{N}~\equiv~\sum_{n=1}^\infty n \, a_n^\dag\cdot a_n~.\eeq{b7}
This expression coincides indeed with its free--string counterpart when it is expressed in terms of the ``$a$"
operators, and as a result the two differ when expressed in terms of the ``$\a$" operators. In the presence of an
EM background, our definition appears more convenient than the one used in~\cite{AN2,AN1,string3}.

\vskip 24pt
\scss{Level $N=1$}

The generic state at this level is
\beq |\Phi\rangle~=~A_\mu(x)\,a^{\dag\mu}_1\,|0\rangle\ . \eeq{b16}
Eq.~(\ref{b13}) is empty, while Eqs.~(\ref{b11}) and (\ref{b12}) reduce to
\bea &\left[\mathcal{D}^2-\tfrac{1}{2}\text{Tr} G^2\right]A_\mu
\, -\, 2\, i\, G_\mu^{~\nu}A_\nu \, =\, 0 \ ,&\label{b17}\\&\mathcal{D}^\mu
\left(\sqrt{\mathbf{1}+iG}\cdot A\right)_\mu \,=\, 0 \ ,&\eea{b18}
on account of the commutation relations~(\ref{r23}) and of the usual definition of the
oscillator vacuum. Letting
\beq \mathcal{A}_\mu\equiv\left(\sqrt{\mathbf{1}+iG}\cdot A\right)_\mu\ ,
\eeq{def1}
these field equations can be cast in the form
\bea &\left[\mathcal{D}^2-
\tfrac{1}{2}\text{Tr}G^2\right]\mathcal{A}_\mu-2iG_\mu^{~\nu}\mathcal{A}_\nu~=~0\ ,&\label{b19}\\
&\mathcal{D}^\mu\mathcal{A}_\mu~=~0\ ,&\eea{b20}
and their algebraic consistency can be easily verified. Notice that in the presence of a
constant EM background the spin-1 field has acquired a new contribution to its mass, $\tfrac{1}{4\alpha'}\text{Tr}G^2$.
The divergence constraint~(\ref{b20}) guarantees that the number of dynamical DoFs is not affected. That the
propagation of the spin-1 field is causal can be shown along the lines of~\cite{AN2}, but we postpone the proof
until Section~\ref{sec:arbitrary}.

\vskip 24pt

\scss{Level $N=2$}

A generic state at this mass level is
\beq |\Phi\rangle~=~h_{\mu\nu}(x)\,a^{\dag\mu}_{1}a^{\dag\nu}_{1}\,
|0\rangle~+~\sqrt{2}\,i\,B_\mu(x)\,a^{\dag\mu}_{2}\,|0\rangle \ , \eeq{b21}
and after the field redefinitions
\bea
\mathcal{H}_{\mu\nu}&\equiv&\left(\sqrt{1+iG}\cdot h\cdot\sqrt{1-iG}\,\right)_{\mu\nu} \ ,\label{def2}\\
\mathcal{B}_\mu&\equiv&\left(\sqrt{1+\tfrac{i}{2}G}\cdot B\right)_\mu \ ,\eea{def3}
one is led to
\bea &\left(\mathcal{D}^2-2-
\tfrac{1}{2}\,\text{Tr}G^2\right)\mathcal{H}_{\mu\nu}-2\,i\,(G\mathcal{H}-\mathcal{H}G)_{\mu\nu}~=~0\ ,&
\label{b25}\\&\left(\mathcal{D}^2-2-\tfrac{1}{2}\,\text{Tr}G^2\right)\mathcal{B}_\mu-2\,i\,G_\mu^
{~\nu}\mathcal{B}_\nu~=~0 \ ,&\label{b26}\\&\mathcal{D}^\mu\mathcal{H}_{\mu\nu}-(1+iG)_\nu^{~\rho}
\mathcal{B}_\rho~=~0\ ,&\label{b27}\\&\mathcal{H}^\mu_{~\mu}+2\mathcal{D}^\mu\mathcal{B}_\mu~=~0 \ .&\eea{b28}

Let us emphasize that, for consistency, in the presence of this non-trivial background the system should have the same
number of dynamical fields as in the free case, and therefore the vector field $\mathcal B_\mu$ should continue
to be non-dynamical. On the other hand, it is simple to see that the system~(\ref{b25})--(\ref{b28}) does not give
rise to any algebraic relation between $\mathcal B_\mu$ and the other fields, while Eq.~(\ref{b26}) contains second
derivatives of $\mathcal B_\mu$. Therefore, $\mathcal B_\mu$ can be non-dynamical only if it is pure gauge.

In order to see that this is actually the case, let us begin by writing the most general \emph{on--shell} gauge
transformation for the system,
\bea \delta\mathcal{H}_{\mu\nu}&=&[J\cdot\mathcal D]_\mu\,\xi_\nu
+[J\cdot\mathcal D]_\nu\,\xi_\mu+\tfrac{1}{2}(K_{\mu\nu}+K_{\nu\mu})(\mathcal D\cdot\xi)\ ,\label{b30}\\
\delta\mathcal{B}_\mu&=&(L\cdot\xi)_\mu+[M\cdot\mathcal D]_\mu(\mathcal D\cdot\xi)\ ,\eea{b31}
where the matrices $J, K, L$, and $M$ are functions of $G$, and the gauge parameter $\xi_\mu$ satisfies
the condition:
\beq \left(\mathcal{D}^2-2-\tfrac{1}{2}\,\text{Tr}G^2\right)\xi_\mu-2\,i\,G_\mu^{~\nu}\xi_\nu~=~0 \ .
\eeq{b29}
To see that Eqs.~(\ref{b30})--(\ref{b31}) define indeed the most general \emph{on--shell} gauge transformation,
let us take $G$ to be small, so that one can restrict the attention to terms up to $\mathcal O(G)$, and let us
examine the possible occurrence of additional higher--derivative terms in Eqs.~(\ref{b30})--(\ref{b31}).
Any such term should be $\mathcal O(G)$, in view of the free limit of the transformations,
Eqs.~(\ref{p27})--(\ref{p28}). It is then simple to realize that, in principle, Eqs.~(\ref{b30}) and~(\ref{b31})
can accommodate terms with odd and even numbers of derivatives, respectively. Moreover, in any such term $G$ should
be contracted with $\xi$\,, since otherwise one could eliminate it using the relation
$G^{\m\n}\mathcal{D}_\m\mathcal{D}_\n\sim\text{Tr}G^2$. One thus finds that
any possible correction of~(\ref{b31}) should contain $\mathcal{D}^2\xi$\,, which however would allow to lower the number
derivatives by two units, due to the on--shell condition~(\ref{b29}). In conclusion, the form of~(\ref{b31}),
with at most two derivatives, is the most general one. Once this is ascertained, looking at the field
equations~(\ref{b25})--(\ref{b28}) one can conclude that no higher derivative
terms can be present in Eq.~(\ref{b30}) if the transformation is to be a symmetry.

One can verify that Eqs.~(\ref{b25})--(\ref{b28}) are invariant under the gauge transformation if
\bea J&=&\mathbf 1 + \tfrac{3}{2}\left(\tfrac{d-6}{d+4}\right)\,i\,G\ ,\label{b32}\\K&=&-\left(\tfrac{10}
{d+4}\right)\,\left[\mathbf 1 + \tfrac{1}{20}\, (d-6)G^2\right]\ ,\label{b33}\\L&=&2\cdot\mathbf 1+
\tfrac{3}{2}\left(\tfrac{d-6}{d+4}\right)\,i\, G\ ,\label{b34}\\M&=&\left(\tfrac{d-6}{d+4}\right)\left[
\mathbf 1+\tfrac{1}{2}\, i\, G\right]\ .\eea{b35}
However, the gauge transformations~(\ref{b30})--(\ref{b29}) are a symmetry \emph{only} in the critical dimension $d=26$.
In this case one can actually gauge away the vector field $\mathcal B_\mu$, ending up with
\beq \left(\mathcal{D}^2-2-\tfrac{1}{2}\,\text{Tr}G^2
\right)\mathcal{H}_{\mu\nu}-2\,i\,(G\mathcal{H}-\mathcal{H}G)_{\mu\nu}~=~0\ ,\qquad \mathcal D^\mu\mathcal
H_{\mu\nu}~=~0\ ,\qquad\mathcal H^\mu_{~\mu}~=~0\ ,\eeq{b37}
so that $\mathcal H_{\mu\nu}$ is a massive spin-2 field, with
\beq
(\text{mass})^2~=~\frac{1}{\alpha'}\, \left(1+\tfrac{1}{4}\,\text{Tr}G^2\right)\ ,
\eeq{mass2}
that possesses a suitable non-minimal coupling to the background EM field. It is manifest that the system~(\ref{b37})
preserves the right number of DoFs, namely $\12(d+1)(d-2)$. That this type of systems admit only causal propagation is shown
in Section~\ref{sec:arbitrary} for the more general case of spin-$s$ tensors, retracing the arguments of Argyres and Nappi~\cite{AN2}.
The Fierz--Pauli system~(\ref{b37})
is indeed related to the Argyres--Nappi Lagrangian of~\cite{AN2}, as we shall see in Section \ref{sec:AN}, but \emph{only}
in $d=26$. In fact, we have already seen here that away from the critical dimension the vector field $\mathcal B_\mu$
cannot be gauged away. This is to be contrasted with the behavior in the absence of the EM background, since we have seen
in the previous section that in the free theory the vector field is pure gauge in any number of space--time dimensions.

To conclude let us remark that, while setting to zero $\mathcal{B}_\mu$ via Eq.~(\ref{b31}) results in a differential equation
in $\xi_\m$, this can still be solved as a power series in $G\mathcal D^2$. One can indeed covariantize the redefinition~(\ref{g1}),
letting
\beq
\mathcal B'_\mu~=~\mathcal B_\mu + \left(\tfrac{d-6}{10}\right)\left[\mathcal D^\nu \mathcal H_{\mu\nu} +\tfrac{1}{8}\left(\tfrac{d+4}{d-1}\right)\mathcal D_\m \mathcal H^\n_{~\n}\right].
\eeq{mp1}
The gauge variation of $\mathcal B'_\mu$ then becomes
\beq
\delta \mathcal B'_\mu~=~L'_{\mu\nu}\xi^{\nu} + \mathcal O(G\mathcal D^2\xi)\ ,
\eeq{shift1}
where $L'_{\m\n}$ is algebraic and invertible for small enough $G$. In order to remove $\mathcal{B}_\mu$,
one is thus led to an iterative definition of $\xi_\m$:
\beq \xi_\mu~=~-{L'}^{-1}_{\mu\nu}\mathcal B'^{\nu} + \mathcal O(G\mathcal D^2 \xi)\ .\eeq{mp2}
\vskip 24pt

\scss{Level $N=3$}

In Section \ref{sec:PSC} we have seen that at this mass level there are two distinct \emph{physical} fields: a
symmetric rank-3 tensor and an antisymmetric rank-2 tensor. The complete gauge fixing of the system, however,
leaves an additional auxiliary scalar field. Our analysis of the $N=2$ level leads one to expect that even in
the presence of a non-trivial background one ought to be able to gauge away unphysical states, at least in the
critical dimension. Given this premise, for the $N=3$ level, one is entitled to begin by considering the state
\beq |\Phi\rangle~=~\phi_{\mu
\nu\rho}(x)\,a^{\dag\mu}_{1}a^{\dag\nu}_{1}a^{\dag\rho}_{1}\,|0\rangle~+~\tfrac{i}{\sqrt2}\,[\,A_{\mu\nu}(x)+\eta
_{\mu\nu}h(x)\,]\,a^{\dag\mu}_{2}a^{\dag\nu}_{1}\,|0\rangle \ , \eeq{b38}
where $\phi_{\mu\nu\rho}$ and $A_{\mu\nu}$ are, respectively, a symmetric 3-tensor and an antisymmetric 2-tensor,
while $h$ is a scalar. When applied to this state, after the field redefinitions
\bea \varPhi_{\mu\nu\rho}&\equiv&(\sqrt{1+iG}\,)^{~\alpha}_\mu(\sqrt{1+iG}\,)_\nu^{~\beta}
(\sqrt{1+iG}\,)_\rho^{~\gamma}\,\phi_{\alpha\beta\gamma}\ ,\label{b42}\\\mathcal{A}_{\mu\nu}&\equiv&
(\sqrt{1+iG}\,)_\mu^{~\alpha}(\sqrt{1+iG}\,)_\nu^{~\beta}\,A_{\alpha\beta}\ ,\eea{b43}
Eqs.~(\ref{b11})--(\ref{b13}) give
\bea &\left(\mathcal{D}^2-4-\tfrac{1}{2}\,\text{Tr}G^2\right)\varPhi_{\mu\nu\rho}+2\,i\,G^\alpha{}_{~(\mu}
\varPhi_{\nu\rho)\alpha}~=~0\ ,&\label{b44}\\&\left(\mathcal{D}^2-4-\tfrac{1}{2}\,\text{Tr}G^2\right)
\mathcal{A}_{\mu\nu}-2\,i\,(G_\mu^{~\alpha}\mathcal{A}_{\alpha\nu}-\mathcal{A}_\mu^{~\alpha}G_{\alpha\nu})~=~0\ ,&
\label{b45}\\&\left(\mathcal{D}^2-4-\tfrac{1}{2}\,\text{Tr}G^2\right)h~=~0\ ,&\label{b45.5}\\
&3\mathcal{D}^\mu\varPhi_{\mu\nu\rho}~=~\tfrac{1}{2}\left[\left\{\sqrt{(1+\tfrac{i}{2}G)(1+iG)}\right\}_\nu^{~~\alpha}
\{\mathcal{A}_{\alpha\rho}+(\sqrt{1+G^2}\,)_{\alpha\rho}\,h\}+(\nu\leftrightarrow\rho)\right]\ ,&\label{b46}\\
&\mathcal{D}^\mu\mathcal{A}_{\mu\nu}~=~\mathcal{D}^\mu\left[(\sqrt{1+G^2}\,)_{\mu\nu}\,h\right]\ ,&\label{b47}\\
&3\varPhi^\mu_{~\mu\nu}+\mathcal{D}_\mu\left[\left(\sqrt{\frac{1+\tfrac{i}{2}G}{1+iG}}\,\right)^{\mu\rho}
\{\mathcal{A}_{\rho\nu}+(\sqrt{1+G^2}\,)_{\rho\nu}\,h\}\right]~=~0\ .&\eea{b48}
As expected, these equations reduce to those of the free theory, Eqs.~(\ref{p43})--(\ref{p44}),
in the limit $G\rightarrow0$, and one can verify their algebraic consistency
making use of the commutation relations \eqref{fancyD2}. One can also conclude
that, just as in the free case, the scalar field $h$ must be auxiliary. In order to see this, let us compute the trace of
Eq.~(\ref{b46}) and let us apply $\mathcal D^\nu$ from the left to Eq.~(\ref{b48}). Subtracting the two resulting
equations and making use of Eq.~(\ref{b47}), one can then obtain
\beq h~=~-\left[\text{Tr}\sqrt{(1+\tfrac{i}
{2}G)(1-iG)}\right]^{-1}\text{Tr}\left[\sqrt{\tfrac{1+\tfrac{i}{2}G}{1+iG}}\cdot\mathcal A\right]\ .\eeq{b49}
Substituting this expression for $h$ into the system~(\ref{b44})--(\ref{b48}), one is finally left with five
independent equations, which are generalizations of Eqs.~(\ref{p46})--(\ref{p47}) in the presence of a constant
EM background, and whose algebraic consistency can be directly verified.

Note that the right--hand side of~(\ref{b49}) does not vanish for $G\neq0$. Therefore, if one were to remove
the scalar $h$ from the generic state~(\ref{b38}), and thus from~(\ref{b44})--(\ref{b48}), the system would be
plagued by algebraic inconsistencies, in the form of an unwarranted constraint on the field $\mathcal A_{\mu\nu}$
that does not exist in the absence of the background. Therefore, any Lagrangian that can consistently describe
the behavior of the two physical fields at level $N=3$ in an EM background must contain an auxiliary scalar mode.
String Theory of course takes care of the problem, since the construction of the spin-3 system includes, in the
first place, a scalar field of this type.

The most important novelty introduced by the background is that the auxiliary scalar is connected, via Eq.~(\ref{b49}),
to the field $\mathcal A_{\mu\nu}$ of the second Regge trajectory. A consistent Lagrangian
description of $\mathcal A_{\mu\nu}$ thus calls for the presence of $h$. On the other hand, $h$ is also the auxiliary
scalar for the spin-3 field of the first Regge trajectory, so that fields belonging to the second Regge trajectory
\emph{cannot} be described consistently without the leading trajectory. At the same time, Lagrangians are bound to
mix the trajectories, and in order to see this in further detail let
us set $\varPhi_{\mu\nu\rho}=0$ in Eqs.~(\ref{b44})--(\ref{b48}). The trace of Eq.~(\ref{b46}) would then give
\beq h~=~-\left[\text{Tr}\sqrt{(1+\tfrac{i}{2}G)(1+iG)
(1+G^2)}\right]^{-1}\text{Tr}\left[\sqrt{(1+\tfrac{i}{2}G)(1+iG)}\cdot\mathcal A\right]\ ,\eeq{b50}
and if Eqs.~(\ref{b49}) and~(\ref{b50}) were to hold one would be led to the unwarranted constraint
\beq G^{\mu\nu}\mathcal A_{\mu\nu}+\mathcal O(G^2)~=~0\ .\eeq{b51}
The system~(\ref{b44})--(\ref{b48}), that involves fields from \emph{both} Regge trajectories, is however algebraically
consistent, and its Lagrangian~\cite{string3} can also be obtained via the BRST method, integrating out auxiliary fields
and performing a complete gauge fixing.

Next, one would like to know whether String Theory can consistently describe, in a constant EM background, a single
spin-3 field of the leading Regge trajectory. With this in mind, let us set to zero the physical field $\mathcal A_{\mu\nu}$
of the subleading trajectory. Interestingly, this choice does not conflict in any way with the algebraic consistency
of the system~(\ref{b44})--(\ref{b48}), which now reduces to the Fierz--Pauli system
\bea &\left(\mathcal{D}^2-4-\tfrac{1}{2}\text{Tr}G^2\right)\varPhi_{\mu\nu\rho}\,+\,2\, i\, G^\alpha{}
_{~(\mu}\varPhi_{\nu\rho)\alpha}~=~0\ ,&\label{b52}\\&\mathcal{D}^\mu\varPhi_{\mu\nu\rho}~=~0\ ,&\label{b53}\\
&\varPhi^\mu_{~\mu\nu}~=~0\ .&\eea{b54}
This is indeed a consistent set of equations, which describes a massive spin-3 field with
\beq
(\text{mass})^2~=~\frac{1}{\alpha'}\, \left(2\,+\, \tfrac{1}{4}\,\text{Tr}G^2\right)\ ,
\eeq{mass3}
and the proper number of propagating DoFs. String Theory also guarantees that this system comes from a Lagrangian,
the Klishevich Lagrangian of~\cite{string3}, with $\mathcal A_{\mu\nu}$ set to zero and after a complete gauge fixing. We have
not shown whether that Lagrangian yields the Fierz--Pauli system away from the critical dimension, but we do not expect it,
since for spin 2 the answer is negative, as we shall see in Section~\ref{sec:AN}.

\vskip 24pt

\scss{Arbitrary mass level: $N=s$}\label{sec:arbitrary}

Let us now focus on the first Regge trajectory, that at this level contains a symmetric rank-$s$ tensor, and let us
investigate the consistency of the string field equations that result from the
physical state conditions \eqref{b11}--\eqref{b13}. To begin with, we thus write the string state
\beq|\Phi\rangle~=~\phi_{\mu_1\mu_2...\mu_s}
(x)\, a^{\dag\mu_1}_{1}a^{\dag\mu_2}_{1}...\ a^{\dag\mu_s}_{1}\,|0\rangle \ ,\eeq{b55}
so that, on account of Eqs.~(\ref{b9.1})--(\ref{b9.3}),
the physical state conditions~(\ref{b11})--(\ref{b13}) give
\bea &\left[\mathcal{D}^2-2(s-1)-\tfrac{1}{2}\,\text{Tr}G^2\right]\phi_{\mu_1...\mu_s}\, +\, 2\, i\, G^\alpha{}_{~(\mu_1}
\phi{}_{\mu_2...\mu_s)\alpha}~=~0\ ,&\label{b56}\\&s\,\mathcal{D}
^\mu\left[(\sqrt{1+iG}\,)_\mu^{~\alpha}\phi_{\alpha\mu_2...\mu_s}\right]~=~0\ ,&\label{b57}\\&\tfrac{1}{2}\, s(s-1)\left[(\sqrt{1+iG}\,)^{\alpha\mu_1}(\sqrt{1+iG}\,)^{~\mu_2}_\alpha
\phi_{\mu_1\mu_2...\mu_s}\right]~=~0\ .&\eea{b58}
It is now convenient to define the symmetric field
\beq \varPhi_{\mu_1\mu_2...\mu_s}~\equiv~(\sqrt{1+iG}\,)_{\mu_1}^{~\alpha_1}(\sqrt{1+iG}\,)_{\mu_2}^{~\alpha_2}
...(\sqrt{1+iG}\,)_{\mu_s}^{~\alpha_s}\phi_{\alpha_1\alpha_2...\alpha_s}\ ,\eeq{b59}
since Eqs.~(\ref{b56})--(\ref{b58}) then give
\bea &\left[\mathcal{D}^2-2(s-1)-\tfrac{1}{2}\,\text{Tr}G^2\right]\varPhi_{\mu_1...\mu_s}+2\,i\,G^\alpha{}_{~(\mu_1}
\varPhi_{\mu_2...\mu_s)\alpha}~=~0\ ,&\label{b60}\\&\mathcal{D}^\mu\varPhi_{\mu\mu_2...\mu_s}~=~0\ ,&\label{b61}\\
&\varPhi^\mu_{~\mu\mu_3...\mu_s}~=~0\ .&\eea{b62}
One can easily show that these equations are algebraically consistent. They form a Fierz--Pauli system for
a massive spin-$s$ field, with a deformed mass, so that now
\beq
(\text{mass})^2~=~\frac{1}{\alpha'}\, \left(s-1+\tfrac{1}{4}\,\text{Tr}G^2\right)\ ,
\eeq{masss}
and it is manifest that the system gives the correct count of DoFs. Eqs.~(\ref{b60})--(\ref{b62}) follow,
at least in $d=26$, from a Lagrangian determined by the BRST method. We have not
derived it, since these relatively simple EoMs suffice for the analysis of the Velo--Zwanziger problem, to which we now turn.

The promised proof of causal propagation for generic spin $s$ can be obtained adapting to our case the arguments
of Argyres and Nappi~\cite{AN2}, and thus resorting to the method of characteristic determinants reviewed briefly
in Appendix~\ref{sec:VZ}. In fact, we have seen that the highest--derivative terms appearing in the EoMs boil down to
the scalar operator $\mathcal D^2$ acting on the fields. From the definition~(\ref{fancyD}) of $\mathcal D_\m$, it is
then clear that the vanishing of the characteristic determinant is tantamount to the condition
\beq \left(G/eF\right)^\m_{~\n}\ n_\m \, n^\n~=~0~,\eeq{c1}
where $G$ is defined in Eq.~\eqref{r13}.

One can perform a Lorentz transformation to reduce $F$ to the block skew--diagonal form\footnote{Leaving aside an exceptional
set of nilpotent fields obeying $F^m=0$ for some $m$.}, $F^{\m}_{~\n}~=~\text{diag}\left(\,F_1~,F_2~,F_3~,...~...\,\right)$\ ,
with the blocks given by
\beq F_1~=~a\left(\begin{array}{cc}
               0 & 1 \\
               1 & 0 \\
             \end{array}
           \right)\ ,\qquad F_{i\neq1}~=~b_i\left(
             \begin{array}{cc}
               0 & 1 \\
               -1 & 0 \\
             \end{array}
           \right)\ ,
\eeq{c3}
where $a$ and $b_i$'s are real--valued functions of the EM field invariants, such that in physically interesting
cases their values are always small. Notice that because of the Lorentzian signature the first block $F_1$ is different
from the $F_{i\neq1}$'s. The same Lorentz transformation will clearly render $G$ block skew--diagonal
as well, $G^{\m}_{~\n}~=~\text{diag}\left(\,G_1~,G_2~,G_3~,...~...\,\right)$\ , with
\beq G_1~=~f(a)\left(
             \begin{array}{cc}
               0 & 1 \\
               1 & 0 \\
             \end{array}
           \right)\ ,\qquad G_{i\neq1}~=~g(b_i)\left(
             \begin{array}{cc}
               0 & 1 \\
               -1 & 0 \\
             \end{array}
           \right)\ ,\eeq{c4}
where
\bea f(a)&\equiv& \frac{1}{\pi}\, [\,\tanh^{-1}(\pi e_0a)+
\tanh^{-1}(\pi e_\pi a)\,]~,\label{c5}\\ g(b_i)&\equiv&\frac{1}{\pi}\, [\,\tan^{-1}(\pi e_0b_i)+
\tan^{-1}(\pi e_\pi b_i)\,]~.\eea{c6}

Let us stress that if the EM field invariants are small, these functions are always well--defined and their absolute
values are much smaller than unity. Given the forms~(\ref{c3}) and (\ref{c4}), one can finally
see that ($G/eF$), where $e=e_0+e_\pi$, is the diagonal matrix
\beq \left(\frac{G}{eF}\right)^\m_{~\n}~=~\text{diag}\left[\,\frac{f(a)}{ea}~,
\frac{f(a)}{ea}~,\frac{g(b_2)}{eb_2}~,\frac{g(b_2)}{eb_2}~,\frac{g(b_3)}{eb_3}~,\frac{g(b_3)}{eb_3}~,
...~...\,\right]~.\eeq{c7}
One can now notice that the functions~(\ref{c5}) and (\ref{c6}) satisfy the inequalities
\beq \frac{f(a)}{ea}~\geq~1~,\qquad 0~<~\frac{g(b_i)}{eb_i}~\leq~1~,\eeq{c8}
so that, in view of~(\ref{c7}), any solution $n_\mu$ of~(\ref{c1}) must be space--like:
\beq n^2~\geq~0~.\eeq{c9}
This is a direct transposition of the $s=2$ argument of \cite{AN2}, and is of course a Lorentz
invariant statement. We can thus conclude that the propagation of the first Regge trajectory is indeed causal,
thanks to the special form of Eqs.~\eqref{b60}--\eqref{b62}, and in particular thanks to the structure of the
non-minimal kinetic terms.

\vskip 36pt

\scs{No--ghost theorem}\label{sec:NG}

The Argyres--Nappi \cite{AN2} and Klishevich \cite{string3} Lagrangians, or for that matter the generalized Fierz--Pauli
conditions of Section \ref{sec:arbitrary}, contain non-standard kinetic contributions, so that it becomes interesting to
investigate whether the flat--space no--ghost theorem extends to this case. We can now show that the no--ghost theorem
(see \emph{e.g.}~\cite{polch}) continues to hold in the regime of physical interest that we identified in the Introduction.

No modifications of the standard arguments are needed in purely magnetic backgrounds, since in this case
the two light--cone coordinates are still subject to standard Neumann boundary conditions. As a result, the Hilbert
space spanned by the $\alpha^\pm_m$ operators maps exactly into that of the free string.

In generic backgrounds where electric fields are also present, matters are more subtle, since the light--cone
directions are affected. However, a no--ghost theorem can still be proved via arguments that follow rather closely those
used for the free string. To this end, it suffices to retrace the proof presented by Polchinski in~\cite{polch},
pp.~139--141. To begin with, one can block--diagonalize the external field strength, a step that can be
carried out for generic constant $F_{\mu\nu}$ backgrounds. When an electric field is present, one can then
modify Eq.~(4.4.7) of~\cite{polch}, turning it into
\beq
[\alpha^{\pm}_m,\alpha^{\mp}_n]~=~-~[m \pm if(a)]~\delta_{m,-n}\ ,\qquad [\alpha^{+}_m,\alpha^{+}_n]~=~
[\alpha^{-}_m,\alpha^{-}_n]~=~0\ ,\eeq{1111}
where $f(a)$ is a skew eigenvalue of $G$, defined in Eq.~(\ref{c5}). When $f(a)=0$, the argument of~\cite{polch}
clearly holds directly, while when $f(a)$ does not vanish one can replace Eq.~(4.4.8) of~\cite{polch} with
\beq N^{\text{lc}}~=~\sum_{m\in\mathbb{Z},\,m\neq0}\,\frac{1}{m - if(a)} \ \alpha^+_{-m}\, \alpha^-_m \ .\eeq{2222}
The next step is to decompose $Q_B$ as in Eq.~(4.4.9) of \cite{polch}, letting  $\alpha^+_0$ play the role of $k^+$.
On the other hand, $Q_1$ and $R$ can be defined exactly as in Eqs.~(4.4.13) and (4.4.14) of \cite{polch}, since $\alpha^-_0$
never appears in their definitions. This also means that a convenient (overcomplete) basis for
the string states is provided by the Fock basis for $\alpha^\pm_{m\neq0}$, together with the coherent states that are
eigenstates of $\alpha^+_0$. One thus finds that the first line of Eq.~(4.4.15) of \cite{polch} should be replaced by
\beq
S~\equiv~\{Q_1,R\}~=~\sum_{m=1}^{\infty}\left\{\,[m+if(a)]\,b_{-m}\,c_m+[m-if(a)]\,c_{-m}\,b_m-\alpha^+_{-m}\,\alpha^-_m
-\alpha^-_{-m}\,\alpha^+_m\,\right\}\ ,
\eeq{3333}
but the rest of the proof carries over verbatim.

\vskip 36pt

\scs{Spin-2 Lagrangians}\label{sec:ANFed}

In the previous sections we have investigated the consistency of our systems at the level of EoMs. While Lagrangians
can be built along the lines of String Theory, they are certainly more complicated than the Fierz--Pauli--like conditions
that we have displayed in Section \ref{sec:arbitrary}. For one matter, as we have seen, they are bound to mix
the leading Regge trajectory with others. In the next subsection
we follow the opposite path, and provide a corollary to \cite{AN2}, showing how their Lagrangian gives rise
to a consistent spin-2 Fierz--Pauli system in the critical dimension $d=26$. In Section~\ref{sec:linAN} we then linearize
the Lagrangian in the EM field strength and suggest a possible field theory program for building consistent Lagrangians
in arbitrary dimensions. We also discuss the gyromagnetic ratio and conclude with a comparative study of the linearized
Argyres--Nappi~\cite{AN2} and Federbush~\cite{fed} Lagrangians.

\vskip 24pt

\scss{Argyres--Nappi Lagrangian and Fierz--Pauli Conditions}\label{sec:AN}

The Argyres--Nappi Lagrangian \cite{AN2} is
\bea L_{\text{AN}}&=&\mathcal{H}_{\mu\nu}^*\left(\mathcal{D}^2
-2-\tfrac{1}{2}\text{Tr}G^2\right)\mathfrak{h}^{\mu\nu}-2i\mathcal{H}_{\mu\nu}^*(G\mathfrak{h}-\mathfrak{h}G)
^{\mu\nu}-\mathcal{H}^*\left(\mathcal{D}^2-2-\tfrac{1}{2}\text{Tr}G^2\right)\mathcal{H}\nonumber\\&&-\,\mathcal{H}_
{\mu\nu}^*\left\{\mathcal D^\mu\mathcal D^\rho[(1+iG)\mathfrak{h}]_\rho^{~\nu}-\tfrac{1}{2}\mathcal D^\mu\mathcal
D^\nu\mathcal H+(\mu\leftrightarrow\nu)\right\}+\mathcal{H}^*\mathcal D^\mu\mathcal D^\nu\mathcal H_{\mu\nu}\ ,\eea{a1}
where $\mathcal D^\mu$ was defined in Eq.~\eqref{fancyD},
\beq \mathcal H_{\mu\nu}\equiv(1+iG)_\mu^{~\alpha}
(1+iG)_\nu^{~\beta}\, \mathfrak{h}_{\alpha\beta}\ ,\eeq{a2}
and for brevity we write $\mathcal{H}$ rather than $\mathcal{H}^\mu_{~\mu}$. One can simply verify that this
Lagrangian is Hermitian and that its variation gives rise to the equations of motion
\bea \mathcal R_{\mu\nu}&\equiv&\left(\mathcal
{D}^2-2-\tfrac{1}{2}\text{Tr}G^2\right)\mathcal{H}_{\mu\nu}-2i(G\mathcal{H}-\mathcal{H}G)_{\mu\nu}-(1+G^2)
_{\mu\nu}\left(\mathcal{D}^2-2-\tfrac{1}{2}\text{Tr}G^2\right)\mathcal H\nonumber\\&&+\tfrac{1}{2}\left\{[(1+iG)
\cdot\mathcal D]_\mu\,[(1+iG)\cdot\mathcal D]_\nu+[(1+iG)\cdot\mathcal D]_\nu\,[(1+iG)\cdot\mathcal D]_\mu\right\}
\mathcal H\nonumber\\&&-\left\{[(1+iG)\cdot\mathcal D]_\mu\,\mathcal D^\rho\mathcal H_{\rho\nu}+[(1+iG)\cdot
\mathcal D]_\nu\,\mathcal D^\rho\mathcal H_{\rho\mu}\right\}+(1+G^2)_{\mu\nu}\mathcal D^\alpha\mathcal D^\beta
\mathcal H_{\alpha\beta}\nonumber\\&=&0~.\eea{a3}
One would like to know whether these equations can be turned into a Fierz--Pauli
system in an \emph{arbitrary} number of space--time dimensions. To this end, let us first take the trace of Eq.~(\ref{a3}),
which gives
\bea \mathcal R^\mu_{~\mu}&\equiv&\left\{(d-2+\text{Tr}G^2)-2iG\right\}^{\alpha\rho}\mathcal D_\rho\mathcal D^\beta
\mathcal H_{\alpha\beta}-\left\{(d-2+\text{Tr}G^2)-G^2\right\}^{\alpha\beta}\mathcal D_\alpha\mathcal D_\beta
\mathcal H\nonumber\\&&+(d-1+\text{Tr}G^2)\left(2+\tfrac{1}{2}\text{Tr}G^2\right)\mathcal H ~=~ 0~.\eea{a4}
On the other hand, the divergence of Eq.~(\ref{a3}) gives
\bea \mathcal D^\mu\mathcal R_{\mu\nu}&\equiv&-\left\{(1+iG)
(2+iG)\right\}_\nu^{~\alpha}\mathcal D^\beta\mathcal H_{\alpha\beta}-\left\{iG(1+iG)\right\}_\nu^{~\alpha}\mathcal D
_\alpha(\mathcal D^\rho\mathcal D^\sigma \mathcal H_{\rho\sigma})\nonumber\\&&+\left\{(1+iG)\left[\left(2-\tfrac{1}{2}
iG+\tfrac{3}{2}G^2\right)+iG\left(\mathcal D^2-\tfrac{1}{2}\text{Tr}G^2\right)\right]\right\}_\nu^{~\alpha}
\mathcal D_\alpha\mathcal H\nonumber\\&=&0\ .\eea{a5}
One can then apply to Eq.~\eqref{a5} the operator $\left[2\mathcal D\cdot(1+iG)
^{-1}\right]^\nu$ from the left to obtain
\bea \left[2\mathcal D\cdot(1+iG)^{-1}\right]^\nu\mathcal D^\mu\mathcal
R_{\mu\nu}&\equiv&\left\{(4-\text{Tr}G^2)-G^2\right\}^{\alpha\beta}\mathcal D_\alpha\mathcal D_\beta\mathcal H+\tfrac{1}{2}
\text{Tr}G^2\left(1+\text{Tr}G^2\right)\mathcal H\nonumber\\
&&-\ \left\{(4-\text{Tr}G^2)-2iG\right\}^{\alpha\rho}\mathcal D_\rho\mathcal D^\beta\mathcal H_{\alpha\beta}
\,=\,0 \ .\eea{a6}
Matters simplify considerably if one adds Eqs.~(\ref{a4}) and (\ref{a6}), obtaining
\beq (d-6+2\text{Tr}G^2)(\mathcal D^\alpha\mathcal D^\beta\mathcal H_
{\alpha\beta}-\mathcal D^2\mathcal H)+\left[2(d-1)+\tfrac{1}{2}\text{Tr}G^2(d+4+2\text{Tr}G^2)\right]\mathcal H=0\ .\eeq{a7}
Finally, applying to Eq.~(\ref{a5}) the operator $\left[\mathcal D\cdot\{(1+iG)(2+iG)\}^{-1}\right]^\nu$
yields
\bea \left[\mathcal D\cdot\{(1+iG)(2+iG)\}^{-1}\right]^\nu\mathcal D^\mu\mathcal R_{\mu\nu}&\equiv& -\left[1+\left(
\frac{iG}{2+iG}\right)^{\mu\nu}\mathcal D_\mu\mathcal D_\nu\right](\mathcal D^\alpha\mathcal D^\beta\mathcal H_{\alpha\beta}
-\mathcal D^2\mathcal H)\nonumber\\&&-\left[\tfrac{1}{4}\text{Tr}G^2+\tfrac{1}{2}(5+\text{Tr}G^2)\left(\frac{iG}{2+iG}
\right)^{\mu\nu}\mathcal D_\mu\mathcal D_\nu\right]\mathcal H\nonumber\\&=&0\ .\eea{a8}

One can now apply the operator
$\left[1+\left\{iG(2+iG)^{-1}\right\}^{\mu\nu}\mathcal D_\mu\mathcal D_\nu\right]$ to Eq.~(\ref{a7}),
multiply Eq.~(\ref{a8}) by $(d-6+2\text{Tr}G^2)$ and add together the results, obtaining
\beq \left[
2(d-1)+\tfrac{1}{4}\,\text{Tr}G^2(d+14+2\,\text{Tr}G^2)-\tfrac{1}{2}(d-26)\left(\frac{iG}{2+iG}\right)^{\mu\nu}
\mathcal D_\mu\mathcal D_\nu\right]\mathcal H~=~0\ .\eeq{a9}
Notice that in an EM background this reduces to an algebraic expression
\emph{only} in the critical space--time dimension $d=26$, whence one obtains the trace constraint
\beq \mathcal H~=~0~\ , \eeq{a10}
so that the situation is quite different from the free case considered in Appendix~\ref{sec:HS}. When~\eqref{a10}
holds,~(\ref{a7}) sets to zero the double divergence of $\mathcal{H}_{\m\n}$, which in its turn yields the divergence
constraint from Eq.~(\ref{a5}). Given the trace and divergence constraints, one can now obtain
\beq
\left(\mathcal{D}^2-2-\tfrac{1}{2}\text{Tr}G^2\right)\mathcal{H}_{\mu\nu}-2i(G\cdot\mathcal{H}
-\mathcal{H}\cdot G)_{\mu\nu}~=~0 \ .
\eeq{a12}
The end result is indeed the deformed Fierz--Pauli system~(\ref{b37}). As we have seen,
this follows from the Argyres--Nappi Lagrangian~(\ref{a1}) \emph{only} in the critical dimension $d=26$.

\vskip 24pt

\scss{Space--time dimensionality and gyromagnetic ratio}\label{sec:linAN}

It is important to notice that the constraint~(\ref{a9}), which would follow from the Argyres--Nappi Lagrangian in an
arbitrary number of dimensions, can be recast in the form
\bea \left[2(d-1)+\tfrac{1}{4}\,\text{Tr}G^2(d+14+2\,\text{Tr}G^2)+\tfrac{1}{2}(d-26)\,
\text{Tr}\left(\frac{G^2}{4+G^2}\right)\right]\mathcal H\nonumber\\-\tfrac{1}{2}(d-26)\left(\frac{G^2}{4+G^2}\right)
^{\mu\nu}\mathcal D_\mu\mathcal D_\nu\mathcal H~=~0 \ ,\eea{l1}
so that for $d\neq26$ it actually fails to be purely algebraic only at $\mathcal O(G^2)$. As a result, if we restrict
ourselves to terms that are at most linear in the EM field strength $F_{\mu\nu}$, the Argyres--Nappi Lagrangian still
gives rise to correct constraints and causal propagation away from the critical dimension. More importantly, the number
of space--time dimensions does not play any role in this case, up to $\mathcal O(F)$. One could thus argue that appropriate
$\mathcal O(F^2)$ terms can always be added, pushing the desired features to $\mathcal O(F^2)$, and so on. While this is
definitely possible when $d=26$, there is no apparent reason why such corrections cannot be added for other dimensions as
well. On top of this, one would need of course correction terms that contain derivatives of $F_{\mu\nu}$ if the latter were
not constant \footnote{This type of construction would have a number of potential applications, which include the improvement
of the holographic model for $d$-wave superconductors of \cite{benini}.}.

Therefore, it becomes interesting to write explicitly the Argyres--Nappi Lagrangian up to terms linear in $F_{\mu\nu}$, restoring
the dependence on $\alpha'$ and regarding $1/\alpha'$ as a generic value $m^2$. The result is
\bea L_{\text{AN}}&=&-|D_\mu\mathfrak{h}_{\nu\rho}|^2+2|D_\mu
\mathfrak{h}^{\mu\nu}|^2+|D_\mu\mathfrak{h}|^2+(\mathfrak{h}^*_{\mu\nu}D^\mu D^\nu \mathfrak{h}+\text{c.c.})-m^2
(\mathfrak{h}^*_{\mu\nu}\mathfrak{h}^{\mu\nu}-\mathfrak{h}^*\mathfrak{h})\nonumber\\&&+\,8ie\text{Tr}(\mathfrak{h}
\cdot F\cdot\mathfrak{h}^*)+\delta L_{\text{kin}}+\mathcal{O}(F^2)\ ,\eea{l2}
where $\delta L_{\text{kin}}$ is a kinetic deformation of $\mathcal{O}(F)$, given by
\bea \delta L_{\text{kin}}&=&-\,i(e/m^2)(F\mathfrak{h}^*-\mathfrak{h}^*F)_{\mu\nu}\left[D^2\mathfrak{h}^{\mu\nu}
-(D^\mu D^\rho\mathfrak{h}_\rho^{~\nu}+D^\nu D^\rho\mathfrak{h}_\rho^{~\mu})+\tfrac12D^{(\mu} D^{\nu)}\mathfrak{h}\right]\nn\\
&&-\,i(e/m^2)D^\m(F\mathfrak{h}^*-\mathfrak{h}^*F)_{\mu\nu}(D_\r\mathfrak{h}^{\r\nu}-D^\n\mathfrak{h})+\text{h.c.} \ . \eea{l3}
As was already mentioned, the Lagrangian~(\ref{l2}) describes consistently a massive spin-2 system coupled to a constant EM
background, up to $\mathcal{O}(F)$.

On the other hand, we note that the spin-2 Federbush Lagrangian of~\cite{fed},
\bea L_{\text{F}}&=&-|D_\mu \varphi_{\nu\rho}|^2+2|D_\mu \varphi^{\mu\nu}|^2+|D_\mu \varphi|^2+
(\varphi^*_{\mu\nu} D^\mu D^\nu \varphi+\text{c.c.})-m^2(\varphi^*_{\mu\nu}\varphi^{\mu\nu}-\varphi^*\varphi)
\nonumber
\\&&+\,ie\text{Tr}(\varphi\cdot F\cdot \varphi^*) \ ,
\eea{v1}
is, up to dimension-4 operators, the only Lagrangian that propagates the correct number of DoFs of a massive spin-2
field in a non-vanishing external EM field $F_{\mu\nu}$. However,
it does not have the same ``dipole" coefficient as the linearized Argyres--Nappi Lagrangian, nor does
it contain, to begin with, dimension-6 kinetic deformations. And indeed, as shown in Appendix~\ref{sec:VZ}, while
Eq.~(\ref{v1}) gives the correct DoF count, it does not take care of hyperbolicity and/or causality.
Therefore, it is interesting to investigate the connection between these two Lagrangians.

At first sight, the connection is simple: up to field redefinitions,
one can conclude that the kinetic deformation present in the Argyres--Nappi Lagrangian is
\beq \delta L_{\text{kin}}~=~-\, 4\, i\, e\, \text{Tr}(\mathfrak{h}
\cdot F\cdot\mathfrak{h}^*) \, +\,  \mathcal{O}(F^2)\ ,
\eeq{deltaAN}
so that to linear order in $F$ the Argyres--Nappi and Federbush
Lagrangians differ only in the coefficient of the dipole term. This coefficient is $4ie$ in the first case,
while it is $ie$ in the second. The first gives a gyromagnetic ratio $g=2$, while the second gives
$g=\12$\,. Intriguingly enough, $g=2$ is a special value that guarantees the absence of high--energy strong
coupling in an important forward ``Compton'' scattering amplitude~\cite{fpt,wb}, and amusingly all open--string
charged states have $g=2$~\cite{fpt}. Upon reflection, this result seems paradoxical, because the Federbush
dipole term is the only one that guarantees the correct number of propagating DoFs, and the number
of DoFs cannot be changed by a local field redefinition! Actually there is no paradox here, simply because the
extra propagating DoF shows up only at $\mathcal{O}(F^2)$~\cite{vz}. Thus, while the linearized Lagrangian is
enough to determine the gyromagnetic ratio implied by the Argyres--Nappi model, it is insufficient to manifest
the subtler problems associated with propagation in external fields. And indeed, expanding the Argyres--Nappi
Lagrangian to $\mathcal{O}(F^2)$ one would find that the kinetic term of the extra DoF is pushed to
higher orders, so that a complete cancelation of the offending mode is guaranteed only by the
full, non-polynomial action.

The existence of a kinetic deformation in the Argyres--Nappi Lagrangian was overlooked in~\cite{deser}, where
it was claimed that no spin-2 Lagrangian propagating the correct number of degrees of freedom could solve the
Velo--Zwanziger problem. This conclusion follows if one assumes from the beginning a canonical
spin-2 kinetic term, and is of course in contradiction with the explicit solution found in \cite{AN2}. Since
the problem with the number of DoFs first manifests itself at $O(F^2)$, it can be solved precisely
via terms like those in Eq.~\eqref{l3}, which are tantamount to a non-derivative Pauli coupling to linear
order in $F$, but which alter the constraint equations to quadratic order.

\vskip 36pt

\scs{Concluding Remarks}\label{sec:conclusion}

The main issue addressed in this paper is whether String Theory can cure the Velo--Zwanziger problem
for a single massive charged spin-$s$ particle in an external EM background. The answer is in the
affirmative, at least for the first Regge trajectory of the open bosonic string, whose symmetric tensors
can be exposed in isolation to constant EM backgrounds. In fact, we showed that
all fields of this type can be described without including other dynamical fields, in that their
generalized Fierz--Pauli conditions
\bea &\left[\mathcal{D}^2\, -\,\frac{1}{\alpha^\prime}\,\left(s-1+\tfrac{1}{4}\,\text{Tr}G^2 \right) \right]
\varPhi_{\mu_1...\mu_s}\,+\, \frac{1}{\alpha^\prime}\,i\,G^\alpha{}_{~(\mu_1}
\varPhi_{\mu_2...\mu_s)\alpha}~=~0\ ,\nonumber\\&\mathcal{D}^\mu\, \varPhi_{\mu\mu_2...\mu_s}~=~0\ ,\label{eqsfinal}\\
&\varPhi^\mu_{~\mu\mu_3...\mu_s}~=~0\ ,\nonumber \eea{generaleq}
are consistent (in an arbitrary number of space--time dimensions)
even in the presence of a constant EM field strength. Moreover, thanks to the special form of their non-minimal kinetic
contributions, these equations result in a causal propagation, thus providing a solution to the Velo--Zwanziger problem
for this class of fields. On the other hand, we have seen in explicit examples that, in general, fields belonging to
subleading trajectories cannot have consistent interactions with an external EM background without additional fields
belonging to other trajectories. Our findings thus resonate with the fact that the Vasiliev systems \cite{vasiliev},
non-linear equations for symmetric tensors of arbitrary rank, can be formulated in an arbitrary number of dimensions.
Conversely, it is natural to regard these systems as an effective description of the first Regge trajectory of the open
bosonic string in a special regime where the remaining excitations decouple.

One may wonder whether the system~(\ref{eqsfinal}) acquires a gauge symmetry when the mass,
$\left(s-1+\tfrac{1}{4}\text{Tr}G^2\right)/\alpha'$, is set to zero. With a finite $\alpha^\prime$, within the regime
of physical interest, this happens only for $s=1$ when $\text{Tr}G^2=0$. On the other hand, for $s>1$ this
would entail the $\alpha^\prime \to \infty$ limit, but then a physically meaningful description would require that
$e F \to 0$, so that $\alpha^\prime e F$ approaches a finite limit. As a result, the higher--spin fields become free
in the limit, consistently with the no--go theorems of~\cite{ww,p}, which state that massless fields with $s>1$ cannot
carry an electric charge.

How unique is the resolution of the original Fierz--Pauli problem that String Theory provides? After all, as was
first noted in~\cite{AN2}, the causality proof (that we retraced in Section~\ref{sec:arbitrary} in order
to extend it to spin-$s$ fields) and
other consistency issues are \emph{not} affected if one makes the replacement $G\rightarrow 2\alpha'eF$.
The complicated function $G$ of the field strength reflects key properties of the string, which can be torn apart
by ``strong" electric fields ($2\pi\alpha^\prime e |\vec E|\sim 1$), and has possibly important lessons in store on the interactions
with non-constant backgrounds~\cite{AN2,string3}. Even slowly varying field strengths, however, are very difficult to
study quantitatively since the string sigma model becomes non-linear in the first place.

String Theory provides a remedy for the Velo--Zwanziger problem but calls for kinetic deformations of the minimal Lagrangian.
It does it in a judicious way, of course: kinetic deformations generically introduce extra DoFs or ghosts,
but the ones present in String Theory do not. While proving this statement is relatively straightforward
-- it is essentially the free--string no--ghost theorem -- we are not aware of any proof to this effect directly in the Lagrangian
theory of massive higher--spin fields. At any rate, non-minimal terms are expected to lower the cutoff of the effective field
theory from that implied by minimal ones, and non-constant external backgrounds would lower it even further.
No simple improvement of the theory thus appears to bypass the upper bound for the cutoff proposed in~\cite{pr3}.

How about the critical dimension, $d=26$? For \emph{free} massive higher spins, it is apparently possible to evade it
rather naturally for low--lying excitations, proceeding from a field theory vantage point, at the price of making some
terms in the Lagrangian or in the (St\"{u}ckelberg) gauge transformations more complicated \cite{RS}. For charged fields
in a constant EM background, if one looks \emph{only} at the EoMs, the dimensionality of space--time does not play any role.
What String Theory guarantees is rather that the EoMs come from a Lagrangian in $d=26$, where the critical dimension is
required in order that the BRST charge be nilpotent. Since nilpotency of the BRST charge is essential in proving the no--ghost
theorem, we do not know if Eqs.~(\ref{eqsfinal}) define a physical, ghost--free system in dimension other than 26.

In some sense, the very appearance of the tensor $G$ may be regarded as evidence
that the underlying theory is inherently non-local, since for instance \emph{two} distinct charges, $e_0$ and $e_\pi$, enter
Eq.~(\ref{r1}), rather than the single charge that a point particle may possess. It is natural to expect that the fully
interacting theory will not be local, and some indications to this effect can be extracted from the limiting behavior of string
amplitudes, as in~\cite{st}.

Finding similar models for massive charged higher--spin fermions starting directly from charged open superstrings (or
from type-0 strings \cite{reviews1,type0}, which contain a plethora of symmetric spinor--tensors) in an
EM background, although possible in principle, seems rather complicated, and apparently no attempts have been made in this
direction. It might be very instructive to look more closely at the indications provided by String Theory for the first few cases,
and in particular for the massive spin-3/2 excitation of the superstring, to see whether they agree with the proposal of~\cite{pr4}.

\vskip 36pt

\section*{Acknowledgments}

We would like to thank F. Benini, A. Campoleoni, D. Francia, C. Iazeolla, E. Joung, M. Taronna and
E. Witten for stimulating discussions and useful comments. MP was supported in part by the NSF grant
PHY-0758032 and by the ERC Advanced Investigator Grant no.\,226455 ``Supersymmetry,
Quantum Gravity and Gauge Fields" (SUPERFIELDS), while RR and AS were supported in part by Scuola Normale
Superiore, by INFN and by the ERC Advanced Investigator Grant no.\,226455 ``Supersymmetry, Quantum Gravity
and Gauge Fields" (SUPERFIELDS).

\vskip 36pt
\begin{appendix}

\scs{Higher--spin systems and EM backgrounds}

In this Appendix we review some basic facts about massive higher--spin fields and their couplings with an
EM background, with special emphasis on some difficulties that are encountered. We refer mostly
to the case of a massive spin-2 field, and mention briefly about symmetric tensors of arbitrary rank.
The reader can find more details and recent results on free higher--spin fields of mixed symmetry
in~\cite{solvay,fms}.

\vskip 12pt

\scss{The Fierz--Pauli $s=2$ system}\label{sec:HS}

Let us begin by reviewing the key properties of the free massive $s=2$ case, which is described in any number
of space--time dimensions by the Fierz--Pauli Lagrangian~\cite{pf}
\beq L_{\text{FP}}=-\tfrac{1}{2}\,(\partial_\mu\vf_{\nu\rho})^2+(\partial_\mu\vf^{\mu\nu})^2 +\tfrac{1}{2}\,
(\partial_\mu\vf)^2\,-\partial_\mu\vf^{\mu\nu}\partial_\nu\vf-\tfrac{1}{2}\, m^2[\vf_{\mu\nu}^2-\vf^2]\ , \eeq{f1}
where for brevity we use the symbol $\vf$ rather than $\vf^\mu_{~\mu}$\ . The corresponding EoMs read
\beq R_{\mu\nu}\,\equiv\,(\Box-m^2)
\vf_{\mu\nu}-\eta_{\mu\nu}(\Box-m^2)\vf+\partial_\mu\partial_\nu\vf-(\partial_\mu\partial^\rho\vf_{\rho\nu}+
\partial_\nu\partial^\rho\vf_{\rho\mu})+\eta_{\mu\nu}\partial^\alpha\partial^\beta\vf_{\alpha\beta}\,=\,0 \ .
\eeq{f2}
Taking divergences and the trace of Eq.~(\ref{f2}) leads to
\bea &\partial^\mu R_{\mu\nu}~\equiv~-m^2
(\partial^\mu\vf_{\mu\nu}-\partial_\nu\vf)~=~0\ ,&\label{f3}\\&\partial^\mu\partial^\nu R_{\mu\nu}~\equiv~
-m^2(\partial^\mu\partial^\nu\vf_{\mu\nu}-\Box\vf)~=~0\ ,&\label{f4}\\&R^\mu_{~\mu}~\equiv~(d-2)(\partial^\mu
\partial^\nu\vf_{\mu\nu}-\Box\vf)+[(d-1)m^2]\,\vf~=~0~,&\eea{f5}
and combining Eqs.~(\ref{f4}) and (\ref{f5}) one arrives at an interesting consequence,
\beq [(d-1)m^2]\,
\vf~=~0\ ,\eeq{f6}
so that for $m^2\neq0$ and $d>1$ one is led to the dynamical trace constraint
\beq \vf~=~0~.\eeq{f7}
The transversality condition follows from Eq.~(\ref{f3}), so that finally Eq.~(\ref{f2})
reduces to the Klein--Gordon equation
\beq (\Box-m^2)\vf_{\mu\nu}~=~0~,\eeq{f9}
which is of course manifestly hyperbolic and causal. Along with the transversality condition,
Eqs.~(\ref{f7})--(\ref{f9}) are the simplest instance of a Fierz--Pauli system, that draws its origin
from the Lagrangian~(\ref{f1}) in an arbitrary number of space--time dimensions, as we have seen.

Trace and divergence conditions are crucial to arrive at the correct number of propagating
DoFs. In $d$ dimensions, a symmetric rank-2 tensor has $\12d(d+1)$ independent components; the
divergence condition eliminates $d$ of them and finally the trace condition removes one more. All in all, one is
thus left with $\12(d+1)(d-2)$ components, which is the correct number of propagating DoFs for a massive spin-2 field.

In general, for a symmetric tensor of arbitrary rank $s$, the Fierz--Pauli system takes the form
\bea \eta^{\mu_1\mu_2}\, \phi_{\mu_1\mu_2 ... \mu_s}&=&0 \ , \label{h1} \\
(\Box-m^2)\, \phi_{\mu_1 ... \mu_s}&=&0 \ , \label{h2} \\
\partial^{\mu_1}\, \phi_{\mu_1 ... \mu_s}&=&0 \ .  \eea{h3}
Its counterpart a for Fermi field, with spin $s=n+\12$\,, contains a $\gamma$-trace condition,
the Dirac equation and a divergence condition:
\bea \gamma^{\mu_1}\, \psi_{\mu_1\mu_2 ... \mu_n}&=&0~,\label{hf1}\\
(\dsl \, - \, m)\, \psi_{\mu_1 ... \mu_n}&=&0~,\label{hf2}\\
\partial^{\mu_1}\, \psi_{\mu_1 ... \mu_n}&=&0~.  \eea{hf3}
%
\scss{Massive $s=2$ field and the Velo--Zwanziger problem}\label{sec:VZ}

One can complexify the spin-2 field in the Lagrangian~(\ref{f1}) and try to minimally couple it to a constant EM
background, following~\cite{fed}. Because covariant derivatives do not commute, the minimal coupling is ambiguous,
so that one is actually led to a family of Lagrangians containing one parameter, which one can call the gyromagnetic
ratio $g$ (see \emph{e.g.}~\cite{deser}):
\bea L&=&-|D_\mu\vf_
{\nu\rho}|^2+2|D_\mu \varphi^{\mu\nu}|^2+|D_\mu \varphi|^2+(\varphi^*_{\mu\nu} D^\mu D^\nu \vf+\text{c.c.})
-m^2(\vf^*_{\mu\nu}\vf^{\mu\nu}-\vf^*\vf)\nonumber\\&&+\,2ieg\,\text{Tr}(\vf\cdot F\cdot\vf^*)~.\eea{vz1}
The resulting EoMs are
\bea 0~~=~~\mathcal R_{\mu\nu}&\equiv&(D^2-m^2)\vf_{\mu\nu}-\eta_{\mu\nu}(D^2-m^2)\vf+\tfrac12
D_{(\mu} D_{\nu)}\vf-\left[D_\m D^\r\vf_{\r\n}+D_\n D^\r\vf_{\r\m}\right]\nonumber\\&&+\,\eta_{\mu\nu}D^\a D^\b
\vf_{\a\b}\,+\, iegF_{\r\m}\vf_{\n}^{~\r}\, +\, iegF_{\r \n}\vf_{\m}^{~\r}~.\eea{vz2}
Combining the trace and the double divergence of Eq.~\eqref{vz2} now gives
\beq \left(\tfrac{d-1}{d-2}\right)m^4\vf~=~ie(2g-1)F^{\m\n}D_\m D^\r\vf_{\r\n}+(g-2)e^2F^{\m\r}F_\r^{~\n}
\vf_{\m\n}-\tfrac34\, e^2F^{\m\n}F_{\m\n}\, \vf~.\eeq{vz6}

The first term on the right--hand side signals a potential DoF breakdown, since a constraint of the free
theory is turned into a propagating field equation unless $g=\12$. The unique minimally coupled model that does not
give rise to a wrong DoF count has therefore $g=\12$, and the
result is precisely the Federbush Lagrangian of~\cite{fed}. With this choice, the divergences and the trace
of Eq.~\eqref{vz2} reduce to
\bea &D^\m\vf_{\m\n}-D_\n\vf~=~\tfrac32
(ie/m^2)\left[F^{\r\s}D_\r\vf_{\s\n}-F_{\n\r}D_\s\vf^{\s\r}+F_{\n\r}D^\r\vf\right]~,&\label{vz7}\\&D^\m D^\n
\vf_{\m\n}-D^2\vf~=~\tfrac32(1/m^2)\left[\,\text{Tr}(F\cdot\vf\cdot F)-\tfrac12\text{Tr}F^2\vf\,\right]~,&
\label{vz8}\\&\vf~=~-\,\tfrac32\left(\tfrac{d-2}{d-1}\right)(e/m^2)^2\left[\,\text{Tr}(F\cdot\vf\cdot F)
-\tfrac12\text{Tr}F^2\vf\,\right]~.&\eea{vz9}
The trace constraint can also be recast in the form
\beq \vf~=~-\,\frac{\tfrac32\left(\tfrac{d-2}{d-1}\right)(e/m^2)^2\,\text{Tr}(F\cdot\vf\cdot F)}{1-\tfrac34\left(\tfrac{d-2}{d-1}\right)(e/m^2)^2\,\text{Tr}F^2} \ ,\eeq{vz10}
an expression that is never singular away from the
instabilities of~\cite{schwinger,nielsen}, \emph{i.e.} in the physically interesting situations where
$\left|\text{Tr}F^2\right|~\ll(m^2/e)^2$. Still, unlike in the free theory,
the trace does not vanish in the presence of the EM background.

However, one still needs to see whether the dynamical DoFs propagate in the correct number
and causally. To this end, let us isolate the terms in Eqs.~(\ref{vz2}) that are of second order
in derivatives,
\beq \mathcal R_{\mu\nu}^{(2)}~=~D^2\vf_{\mu\nu}-\left[D_\m(D^\r\vf_{\r\n}-D_\n\vf)+
(\m\leftrightarrow\n)\right]-\tfrac12 D_{(\mu} D_{\nu)}\vf+\eta_{\mu\nu}(D^\a D^\b\vf_{\a\b}-D^2\vf)~,\eeq{vz11}
where the last can be actually dropped in view of~(\ref{vz8}) while the constraint equations~(\ref{vz7}) and~(\ref{vz9})
can be substituted in the second and third terms. The end result,
\bea \mathcal R_{\mu\nu}^{(2)}&=&\Box\vf_
{\m\n}-\tfrac32(ie/m^2)\left[F^{\r\s}\de_\r\de_{(\m}\vf_{\n)\s}+F_{\r(\m}\de_{\n)}(\de_\s\vf^{\s\r}-\de^\r\vf)\right]
\nn\\&&+\,\tfrac32\left(\tfrac{d-2}{d-1}\right)(e/m^2)^2\left[F^{\r\s}F_\s^{~\l}\de_\m\de_\n\vf_{\r\l}-\tfrac12
\text{Tr}F^2\de_\m\de_\n\vf\right]\ ,\eea{vz12}
is the counterpart, for the model, of the Klein--Gordon equation.

Following \cite{vz}, one can now resort to the characteristic determinant method to investigate the causal properties of
the system, replacing $i\partial_\mu$ with $n_\mu$, the normal to the characteristic hypersurfaces, in the highest--derivative
terms of the EoMs. The determinant $\Delta(n)$ of the resulting coefficient matrix determines in fact the causal
properties of the system, and in particular if the algebraic equation $\Delta(n)=0$ has real solutions for $n_0$ for any
$\vec{n}$, the system is hyperbolic, with maximum wave speed $n_0/|\vec{n}|$. On the other hand, if there are time--like
solutions $n_\mu$ for $\Delta(n)=0$, the system admits acausal propagation. Note that the procedure is akin to solving the
EoMs in the eikonal approximation, letting
\beq \vf_{\m\n}~=~\hat{\vf}_{\m\n}\exp(itn\cdot x)\ ,\qquad t\rightarrow\infty\ .\eeq{eikonal}

The coefficient matrix determined by (\ref{vz12}) takes the form
\bea M_{(\m\n)}^{~~(\a\b)}(n)&=&-\tfrac12n^2
\delta^{(\a}_\m\delta^{\b)}_\n+\tfrac34(ie/m^2)\left[n_\r F^{\r(\a}n_{(\m}\delta_{\n)}^{\b)}-n_{(\m}F_{\n)}^{~(\a}n^{\b)}
+2n_{(\m}F_{\n)}^{~\r}n_{\r}\eta^{\a\b}\right]\nn\\&&-\,\tfrac32\left(\tfrac{d-2}{d-1}\right)(e/m^2)^2\,n_\m n_\n\left[F^{\a\r}F_\r^{~\b}-\tfrac12\text{Tr}
F^2\eta^{\a\b}\right]~.\eea{vz13}
This expression should indeed be regarded as a matrix whose $\tfrac12d(d+1)$ rows and columns are labeled by pairs
of Lorentz indices $(\m\n)$ and $(\a\b)$. In particular in four dimensions its determinant reads
\beq \Delta(n)~=~(n^2)^8\left[n^2\, -\, \left(\tfrac{e}{m^{2}}\right)^2\, \left(\tilde{F}\cdot n\right)^2\right]
\left[n^2\, +\, \left(\tfrac{3e}{2m^{2}}\right)^2 \, \left(\tilde{F}\cdot n\right)^2\right]~,\eeq{vz14}
where $\tilde{F}_{\m\n}\equiv\tfrac12\e_{\m\n\r\s}F^{\r\s}$, so that
\beq
(\tilde{F}\cdot n)^2~\equiv~(n_0\vec B+\vec
n\times\vec E)^2-(\vec n\cdot\vec B)^2
\ .\eeq{formula}

Let us now consider four--dimensional EM invariants such that $\vec B\cdot\vec E=0$, $\vec B^2-\vec E^2>0$,
which entails that $\vec B^2$ is non-vanishing in all Lorentz frames. One can always find a vector $\vec n$,
perpendicular to $\vec B$, for which the characteristic determinant vanishes if
\beq \frac{n_0}{|\vec n|}~=~\frac{1}{\sqrt{1-\left(\tfrac{3e}{2m^{2}}\right)^2\vec B^2}}~,\eeq{vz15}
thanks to the last factor appearing in~(\ref{vz14}).
As a result, superluminal propagation can occur even for infinitesimally small values of $\vec B^2$, and the
propagation itself ceases to occur whenever
$\vec B^2\geq\left(\tfrac{2}{3}m^2/e\right)^2$. Actually, one can always find a Lorentz frame where the pathology
shows up, and in particular the magnetic field $\vec B^2$ can reach the critical value in
a frame where $\vec E^2=\left(\tfrac{2}{3}m^2/e\right)^2-\e$, with $\e$ arbitrarily small.
This is the most serious aspect of the problem: it persists even for very small values of the EM field invariants,
far away from the instabilities of \cite{schwinger,nielsen}, where one would expect to be dealing with well--behaved
and long--lived propagating particles. This is the so--called Velo--Zwanziger problem~\cite{vz}, which as we have just
recalled arises already at the classical level \footnote{
The pathology in the corresponding quantum mechanical theory, for $s=3/2$, was found much earlier by Johnson and
Sudarshan~\cite{js}. From a canonical viewpoint, the equal time commutation
relations become ill--defined in an EM background. That the Johnson--Sudarshan and Velo--Zwanziger problems have a common
origin was later shown in~\cite{same}.}. It is important to note that the pathology is not a special property of the spin-2 case,
but it is expected to persist for all charged massive particles with $s\geq3/2$, since it originates from the very
existence of longitudinal modes of massive high--spin particles~\cite{pr1}.

\vskip 24pt

\scs{The bosonic string: notation and conventions}\label{sec:states}

In this Appendix we spell out a few basic facts about the open bosonic string that have some
bearing on our derivations. As in the main body of the paper we work with
signature $(-,+)$ on the world sheet
and in units such that $\alpha^\prime = \frac{1}{2}$. The standard open strings with Neumann
boundary conditions are then described by the coordinate functions
\beq X^\mu_{\text{free}}(\tau,\sigma)~=~x^\mu\, +\, \tau\, \bar{\alpha}_0^\mu\, +\, \frac{i}{2}\sum_{m\neq0}
\frac{1}{m}\left[\text{e}^{-im(\tau+\sigma)}\, +\, \text{e}^{-im(\tau-\sigma)}\right]\,\bar{\alpha}_m^\mu\ ,\eeq{r7}
where $x^\mu$ is the center--of--mass coordinate while the term proportional to $\tau$
is the corresponding momentum. The mode functions are thus, aside from a constant,
\bea \bar{\Psi}_m(\tau,\sigma)&=&\tfrac{i/2}{\sqrt{m}}\,\text{e}^{-im\tau}[\,\text{e}^{-im\sigma}
+\text{e}^{im\sigma}\,]\qquad m\in\mathbb{N}_1 \ , \label{r8}\\\bar{\Psi}_0(\tau,\sigma)&=&\tau \ , \eea{r9}
which obey orthogonality relations that can be presented in the convenient form
\beq (\bar{\Psi}_m,\bar{\Psi}_n)~\equiv~\frac{1}{\pi}\int_0^\pi
d\sigma \bar{\Psi}_m^*(\tau,\sigma)\star\bar{\Psi}_n(\tau,\sigma)~=~\delta_{mn}(1-\delta_{m0})
\qquad m,n\in\mathbb{N}_0\ ,\eeq{r10}
where $\star\equiv i\overleftrightarrow{\partial_\tau}=i
\overrightarrow{\partial_\tau}-i\overleftarrow{\partial_\tau}$. The constant mode, just like
$\bar{\Psi}_0$, is orthogonal to all $\bar{\Psi}_{m\in\mathbb{N}_1}$ and has a vanishing norm for
the inner product~(\ref{r10}), but is not orthogonal to $\bar{\Psi}_0$
\beq (1,1)~=~0\ ,\qquad
(1,\bar{\Psi}_m)~=~i\delta_{m0}\ .\eeq{r11}
The mode functions split naturally into two mutually orthogonal subsets, particle--like $\{1,\bar{\Psi}_0\}$,
and string--like $\{\bar{\Psi}_{m\in\mathbb{N}_1}\}$, and the infinitely many string--like modes form an orthonormal
set of functions. For the free string their orthonormality relation is usually presented in a more familiar form
that does not involve $\tau$, but this form extends naturally to the case of a constant EM
background, as reviewed in Section~\ref{sec:mode}. The two particle--like modes have zero norm and a
non-vanishing mutual inner product, so that
\beq x^\mu~=~ i(\bar{\Psi}_0, X^\mu)\ ,\qquad p^\mu~=~-i(1, X^\mu)\ .\eeq{r11.5}

The free string modes $\bar{\alpha}_m$ in~(\ref{r7}) obey the commutation relations
\beq [\,\bar{\alpha}_m^\mu,\bar{\alpha}_n^\nu\,]~=~m\eta^{\mu\nu}\delta_{m,-n}\qquad m,n\in\mathbb{Z}~,\eeq{p1}
that correspond to the $G\rightarrow0$ limit of Eq.~(\ref{r22}).
The oscillators $\bar{\alpha}_{m\neq0}$ define an infinite set of creation and annihilation
operators,
\beq a_m^\mu~=~\tfrac{1}{\sqrt{m}}\,\bar{\alpha}_m^\mu\ ,\qquad a_m^{\dag\mu}~=~\tfrac{1}{\sqrt{m}}\,
\bar{\alpha}_{-m}^\mu \qquad m\in\mathbb{N}_1\ ,\eeq{p3}
so that one can consider general linear combinations of terms obtained applying creation
operators to the ground state:
\beq |\Psi\rangle~=~ \sum_{s=1}^\infty\sum_{m_i=1}^\infty\psi_{\mu_1\mu_2...\mu_s}
^{(m_1m_2...m_s)}\,a_{m_1}^{\dag\mu_1}a_{m_2}^{\dag\mu_2}\,...\,a_{m_s}^{\dag\mu_s}\,|0\rangle\ .\eeq{p6}
Given a set of integers $(m_1,m_2,...,m_s)$,
the coefficient function $\psi_{\mu_1\mu_2...\mu_s}^{(m_1m_2...m_s)}$ is a rank-$s$ Lorentz tensor, generically of
mixed symmetry, that is interpreted as a field associated to the corresponding string state, and as such is a
function of the string center--of--mass coordinates.

Of particular interest are the string states that are eigenstates of the number operator,
\beq \mathcal{N}~\equiv~\sum_{n=1}^\infty n\, a_{n}^\dag\cdot a_n\ ,\eeq{p7}
whose integer eigenvalues $N$ determine the masses of the open--string states according to
\beq (\text{mass})^2~=~2(N-1)~=~\frac{1}{\alpha^{\prime}}\,(N-1)~,\eeq{p15}
where in the last step we have reinstated $\alpha^\prime$.

A ``physical'' string state is to satisfy some conditions that involve the Virasoro generators
\beq L_n~=~\tfrac{1}{2}\sum_{m\in\mathbb{Z}}\bar
{\alpha}_{n-m}\cdot\bar{\alpha}_{m} \qquad n\in\mathbb{N}_0\ .\eeq{p9}
$L_0$, $L_1$, and $L_2$ are particularly important, since all other generators with $n\in\mathbb{N}_0$ can be
recovered from their commutators. In terms of the ``$a$" operators they read
\bea L_0&=&-\tfrac{1}{2}\,\Box+\sum_{m=1}^{\infty}m\,a_m^\dag
\cdot a_m~\equiv~-\tfrac{1}{2}\,\Box+\mathcal{N}\ ,\label{p9.1}\\L_1&=&-\, i\, \partial\cdot a_1\, +\, \sum_{m=2}^
{\infty}\sqrt{m(m-1)}\,a_{m-1}^\dag\cdot a_m\ ,\label{p9.2}\\L_2&=&-\sqrt{2}\, i\, \partial\cdot a_2\, +\, \tfrac{1}{2}\,
a_1\cdot a_1+\sum_{m=3}^{\infty}\sqrt{m(m-2)}\,a_{m-2}^\dag\cdot a_m\ .\eea{p9.3}
As we have pointed out, demanding that $L_0$, $L_1$ and $L_2$ annihilate a physical state translates precisely,
in terms of the coefficient fields, into the Fierz--Pauli conditions \eqref{h1}--\eqref{h3}.

\vskip 24pt

\end{appendix}

\end{document}